\documentstyle[aps,prb,epsfig]{revtex}
\begin{document}
\draft
\author{Mikio Eto$^{1,2}$ and Yuli V.\ Nazarov$^{1}$}
\address{$^1$Department of Applied Physics/DIMES,
Delft University of Technology, \\
Lorentzweg 1, 2628 CJ Delft, The Netherlands \\
$^2$Faculty of Science and Technology, Keio
University, \\
3-14-1 Hiyoshi, Kohoku-ku, Yokohama 223-8522, Japan}
\title{Mean Field Theory of the Kondo Effect in Quantum Dots with
an Even Number of Electrons}
\date{January 9, 2001}
\maketitle
\begin{abstract}
We investigate the enhancement of the Kondo effect in quantum dots with
an even number of electrons, using a scaling method and
a mean field theory.
We evaluate the Kondo temperature $T_{\rm K}$ as a function of the
energy difference between spin-singlet and triplet states in the dot,
$\Delta$, and the Zeeman splitting, $E_{\rm Z}$.
If the Zeeman splitting is small, $E_{\rm Z} \ll T_{\rm K}$,
the competition between the singlet and triplet states enhances the
Kondo effect. $T_{\rm K}$ reaches its maximum around $\Delta=0$
and decreases with $\Delta$ obeying a power law.
If the Zeeman splitting is strong, $E_{\rm Z} \gg T_{\rm K}$,
the Kondo effect originates from the degeneracy between the singlet
state and one of the components of the triplet state at
$-\Delta \sim E_{\rm Z}$. We show that $T_{\rm K}$ exhibits another power-law
dependence on $E_{\rm Z}$. The mean field theory provides
a unified picture to illustrate the crossover between these regimes.
The enhancement of the Kondo effect can be understood in terms of
the overlap between the Kondo resonant states created around the
Fermi level. These resonant states provide
the unitary limit of the conductance $G\sim 2e^2/h$.
\end{abstract}

\section{Introduction}
 
The Kondo effect observed in semiconductor quantum dots
has attracted a lot of interest.\cite{Kondo1,Kondo2,Kondo3,Kondo4,Wilfred}
In a quantum dot, the number of electrons $N$ is fixed by
the Coulomb blockade to integer values and can be tuned by the gate
voltage. Usually the discrete spin-degenerate levels in the quantum dot
are consecutively occupied, and the total spin is zero or 1/2
for an even and odd number of electrons, respectively.
The Kondo effect takes place only in the latter case.
The spin $1/2$ in the dot is coupled to the Fermi sea
in external leads through tunnel barriers, which
results in the formation of the Kondo resonant state at the Fermi
level.\cite{classics,classic2,classic3}
The conductance through the dot is enhanced to a value of the order of
$e^2/h$ at low temperatures of $T \ll T_{\rm K}$
(Kondo temperature).\cite{Glazman,Ng,Kawabata,Hershfield,Meir}
This is called unitary limit. When $N$ is even, there is no localized
spin and thus the Kondo effect is not relevant.

Recently Sasaki {\it et al}.\ has found a large Kondo effect
in so-called ``vertical'' quantum dots with an even $N$.\cite{Sasaki}
The spacing of discrete levels in such dots is comparable with
the strength of electron-electron Coulomb interaction. Hence
the electronic states deviate from the simple picture mentioned
above.\cite{Tarucha,Leo}
If two electrons are put into nearly degenerate levels,
the exchange interaction favors a spin triplet
(Hund's rule).\cite{Tarucha} This state is changed to a
spin singlet by applying a magnetic field which
increases the level spacing. Hence the energy difference between the
singlet and triplet states, $\Delta$, can be controlled experimentally
by the magnetic field. The Kondo effect is significantly enhanced around
the degeneracy point between the triplet and singlet states, $\Delta=0$.
Tuning of the energy difference between the spin states is hardly
possible in traditional Kondo systems of dilute magnetic impurities in
metal and thus this situation is quite unique to the quantum dot systems.

The Kondo effect in multilevel quantum dots has been investigated
theoretically by several groups.\cite{Inoshita,Pohjola,Izumida,Yeyati}
They have shown that the contribution from multilevels enhances
the Kondo effect.
In our previous paper,\cite{me} we have considered the experimental
situation by Sasaki {\it et al}.\ in which
the spin-singlet and triplet states are almost degenerate.
We have calculated the Kondo temperature $T_{\rm K}$
as a function of $\Delta$,
using the poor man's scaling method.\cite{Anderson,multiK,multiK2}
We have shown that $T_{\rm K}(\Delta)$ is maximal around $\Delta=0$
and decreases with increasing $\Delta$ obeying a power law,
$T_{\rm K}(\Delta) \propto 1/\Delta^{\gamma}$. The exponent $\gamma$ is not
universal but depends on a ratio of the initial coupling constants.
Our results indicate that the Kondo effect is enhanced by
the competition between singlet and triplet states, in agreement with
the experimental findings.\cite{Sasaki}

We have disregarded the Zeeman splitting of the spin-triplet state,
$-E_{\rm Z} M$ ($M=0,\pm 1$ is $z$-component of the total spin $S=1$),
since this is a small energy scale in the
experimental situation, $E_{\rm Z} \ll T_{\rm K}$.\cite{Sasaki}
Pustilnik {\it et al}.\ and Giuliano {\it et al}.\
have studied another situation where the Zeeman effect is relevant,
$E_{\rm Z} \gg T_{\rm K}$.\cite{Pustilnik,Giuliano} Usually the Zeeman
effect lifts off the degeneracy of the spin states and, as a result,
breaks the Kondo effect. They have found that the Kondo effect can
arise from extra degeneracy between one of the components of the spin-triplet
state, $|S M \rangle=|1 1 \rangle$, and a singlet state, $|0 0 \rangle$,
if the value of $\Delta$ is tuned to fulfill that $E_{\rm Z}=-\Delta$.
Their mechanism might explain some other experimental results of
the Kondo effect in quantum dots under high magnetic
fields.\cite{Kondo4,Leo2}

The purpose of the present paper is to construct a general theory for
the enhancement of
the Kondo effect in quantum dots with an even number of electrons,
for various values of $\Delta$ and $E_{\rm Z}$.
We adopt the poor man's scaling method along with the mean field
theory. It is well known that
the characteristic energy scale of the Kondo physics, the Kondo
temperature $T_{\rm K}$, is determined by all the energies from $T_{\rm K}$ up
to the upper cutoff.\cite{classic2,classic3}
By the scaling method, we can evaluate $T_{\rm K}$ (its exponential part
at least) by taking all the energies properly.\cite{Anderson,multiK,multiK2}
When $E_{\rm Z}$ is negligible, the energies from $\Delta$ to the upper
cutoff would feel fourfold degeneracy of the dot states, 
$|1 M \rangle$ ($M=0,\pm 1$) and $|0 0 \rangle$, which enhances
the Kondo temperature. With increasing $\Delta$, $T_{\rm K}$ decreases by
a power law.\cite{me}
We extend our previous calculations to the case of
$E_{\rm Z}=-\Delta \gg T_{\rm K}$
which has been discussed by Pustilnik {\it et al}.\cite{Pustilnik}
and Giuliano {\it et al}.\cite{Giuliano}
We take into account the energies not only from $T_{\rm K}$ to $E_{\rm Z}$,
where only two degenerate states $|1 1 \rangle$ and $|0 0 \rangle$ are
relevant, but also from $E_{\rm Z}$ to the upper cutoff, where the
dot states seem fourfold degenerate. The latter energy region has been
neglected in Refs.\ \onlinecite{Pustilnik} and \onlinecite{Giuliano}.
In consequence we find a power law dependence of $T_{\rm K}$ on
$E_{\rm Z}$ again.

The mean field theory of the Kondo effect was pioneered by
Yoshimori and Sakurai\cite{Yoshimori} and is commonly used for the
Kondo lattice model.\cite{Lacroix}
It is useful to capture main qualitative features of the Kondo
effect; renormalizability at the scale of $T_{\rm K}$, resonances at the Fermi
level, and resonant transmission.
The simplicity and universality of the mean field theory
have driven us to apply it to the problem in question.
Generally the Kondo effect gives rise to a many-body ground state which
consists of the dot states $|SM \rangle=f_{SM}^{\dagger}|0\rangle$
and the conduction electrons $\Pi c_{k\sigma}^{\dagger} |0 \rangle$.
The total spin of this ground state is less than the original spin $S$ 
localized in the dot. The binding energy is of the order of
the Kondo temperature $T_{\rm K}$.
We take into account the spin couplings
between the dot states and conduction electrons,
$\langle f_{SM}^{\dagger}c_{k\sigma} \rangle$,
by the mean field, neglecting their fluctuations.\cite{com0}
These spin couplings give rise to resonant
states around the Fermi level $\mu$ with the width of
the order of $T_{\rm K}$.
The conduction electrons can be transported through
the resonant levels, which yields the unitary limit of the
conductance $G \sim 2e^2/h$.
For our study, the mean field calculations have the following advantages.
(i) The enhancement of $T_{\rm K}$ by the competition between the singlet
and triplet states can be directly understood in terms of the overlap
between their Kondo resonant states. (ii) The power law dependence of
$T_{\rm K}$ on $\Delta$ or $E_{\rm Z}$ is obtained, which is in accordance
with the calculated results by the scaling method.
(iii) The mean field calculations are applicable to the
intermediate regions where two of $T_{\rm K}$, $\Delta$, and $E_{\rm Z}$,
are of the same order. The poor man's scaling method
hardly gives any results
in these regions. Hence we can examine the whole parameter region of
$\Delta$ and $E_{\rm Z}$ by the mean field theory.
The disadvantage of the mean field calculations is
that they only give qualitative answers.\cite{com0}
Hence the mean field theory and scaling method are
complementary to each other for understanding the Kondo effect.

We shall discuss the relation of our approach to the renormalization group
analysis of the multilevel Kondo effect.\cite{multiK,multiK2}
Our model effectively reduces to the one with two channels in the leads and
spin-triplet (and singlet) state in the dot. The ground state of this model
would be believed to be a spin singlet, which corresponds to the full screening
of the dot spin. The poor man's scaling approach and our mean field
theory, however, show a tendency to the formation of the underscreened
Kondo ground state with spin $1/2$. We should mention that the exact
ground state can not be determined within the limits of the
applicability of these approaches. Pustilnik and Glazman have recently
proposed a different model for the ``triplet-singlet Kondo
effect.''\cite{Pustilnik2} In our notations, they set $C_1=\sqrt{2}, C_2=0$
in Eq.\ (\ref{eq:singlet}) for the singlet state.
Their model can be directly mapped onto a special case of the two-impurity
Kondo model,\cite{Ludwig} for which the ground state is a spin singlet.
We are concerned about the case of $C_1 \approx C_2$, and we find that
the difference between $C_1$ and $C_2$ reduces as a result of
the renormalization.\cite{me} This suggests that the case considered
in Ref.\ \onlinecite{Pustilnik2} is by no means a generic one.

This paper is organized as follows.
Our model is presented in the next section. In section III, we rederive
$T_{\rm K}(\Delta)$ when the Zeeman splitting is irrelevant,
using the poor man's scaling method, in a simpler form than our previous
work.\cite{me} Then we extend our calculations to the
case of $E_{\rm Z}=-\Delta \gg T_{\rm K}$. The section IV is devoted to the
mean field theory for the Kondo effect in quantum dots. First we
explain this theory for the usual Kondo effect in a quantum dot with
$S=1/2$. Then we apply the mean field scheme
to our model with an even number of electrons in the dot.
The conclusions and discussion are given in the last section.

\section{Model}

We are interested in the competition between the spin-singlet and
triplet states in a quantum dot.
To model the situation, it is sufficient to consider two extra electrons
in a quantum dot at the background of a singlet state of all other
$N-2$ electrons, which we will regard as the vacuum $|0 \rangle$.
These two extra electrons occupy two levels of different orbital
symmetry.\cite{symmetry} The energies of the levels are
$\varepsilon_1$ and $\varepsilon_2$.
Possible two-electron states are (i) the spin-triplet state,
(ii) the spin-singlet state of the same orbital symmetry as
the triplet state,
$1/\sqrt{2}(d_{1 \uparrow}^{\dagger} d_{2 \downarrow}^{\dagger}
 -d_{1 \downarrow}^{\dagger} d_{2 \uparrow}^{\dagger}) |0 \rangle$,
and (iii) two singlet states of different orbital symmetry,
$d_{1 \uparrow}^{\dagger} d_{1 \downarrow}^{\dagger} |0 \rangle$,
 $d_{2 \uparrow}^{\dagger} d_{2 \downarrow}^{\dagger} |0 \rangle$.
Among the singlet states,
we only consider a state of the lowest energy, which belongs
to the group (iii).
Thus we restrict our attention to four
states, $|S M \rangle$: 
\begin{eqnarray}
|1 1 \rangle & = & d_{1 \uparrow}^{\dagger} d_{2 \uparrow}^{\dagger}
|0 \rangle  \\
|1 0 \rangle & = & \frac{1}{\sqrt{2}} (d_{1 \uparrow}^{\dagger} d_{2
\downarrow}^{\dagger}
                        +d_{1 \downarrow}^{\dagger} d_{2
\uparrow}^{\dagger}) |0 \rangle \\
|1 -1 \rangle & = & d_{1 \downarrow}^{\dagger} d_{2 \downarrow}^{\dagger}
|0 \rangle \\
|0 0 \rangle & = & \frac{1}{\sqrt{2}} (C_1 d_{1 \uparrow}^{\dagger}
d_{1 \downarrow}^{\dagger}
 -C_2 d_{2 \uparrow}^{\dagger} d_{2 \downarrow}^{\dagger}) |0 \rangle,
\label{eq:singlet}
\end{eqnarray}
where $d_{i \sigma}^{\dagger}$ creates an electron with spin $\sigma$ in
level $i$.
The coefficients in the singlet state, $C_1$, $C_2$ ($|C_1|^2+|C_2|^2=2$),
are determined by the electron-electron interaction
and one-electron level spacing $\delta=\varepsilon_2-\varepsilon_1$.
We set $C_1=C_2=1$. This is the case for $\delta=0$.\cite{com1}
Although $C_1 \ne C_2$ in general,
the scaling analysis shows that the Kondo temperature is the same as
that in the case of $C_1=C_2=1$, apart from a prefactor.\cite{me}
The energies of the triplet state are given by
\begin{equation}
E_{S=1,M}=E_{S=1}-E_{\rm Z} M
\end{equation}
and the energy of the singlet state is denoted by
$E_{00}$. We define $\Delta$ by
\begin{equation}
\Delta=E_{00}-E_{S=1}.
\end{equation}
The energy diagram for the spin states is indicated in Fig.\ 1(a).

The dot is connected to two external leads $L$, $R$ with free electrons
being described by
\[
H_{\rm{leads}}=\sum_{\alpha=L,R}\sum_{k \sigma i}
\varepsilon_{k}^{(i)} c_{\alpha, k\sigma}^{(i) \dagger}
c_{\alpha, k\sigma}^{(i)},
\]
where $c_{\alpha, k\sigma}^{(i) \dagger}$ ($c_{\alpha, k\sigma}^{(i)}$)
is the creation (annihilation) operator of
an electron in lead $\alpha$ with momentum $k$, spin $\sigma$, and
orbital symmetry $i$ $(=1,2)$. 
The density of states $\nu$ in the leads remains constant in the
energy band of $[-D, D]$.
The tunneling between the dot and the leads is written as
\[
H_T=\sum_{\alpha=L,R} \sum_{k \sigma i} (V_{\alpha,i}c_{\alpha,
k\sigma}^{(i) \dagger} d_{i \sigma} + \rm{H.c.}).
\]
We assume that the orbital symmetry is conserved in the tunneling
processes.\cite{symmetry}
To avoid the complication due to the fact that there are two leads
$\alpha=L,R$, we perform a unitary
transformation for electron modes in the leads along the lines of
Ref.\ \onlinecite{Glazman};
$c_{k \sigma}^{(i)}=(V_{L,i}^*c_{L,k \sigma}^{(i)}+
V_{R,i}^*c_{R,k \sigma}^{(i)})/V_i$,
$\bar{c}_{k \sigma}^{(i)}=(-V_{R,i}c_{L,k \sigma}^{(i)}+
V_{L,i}c_{R,k \sigma}^{(i)})/V_i$,
with $V_i=\sqrt{|V_{L,i}|^2+|V_{R,i}|^2}$. The modes
$\bar{c}_{k \sigma}^{(i)}$ are not coupled to the quantum dot and
shall be disregarded hereafter.
Then $H_{\rm{leads}}$ and $H_T$ are rewritten as
\begin{eqnarray}
H_{\rm{leads}} & = & \sum_{k \sigma i}
\varepsilon_{k}^{(i)} c_{k\sigma}^{(i) \dagger}
c_{k\sigma}^{(i)}, \\
H_T & = & \sum_{k \sigma i} V_{i}
(c_{k\sigma}^{(i) \dagger} d_{i \sigma} + \rm{H.c.}).
\end{eqnarray}

We assume that the state of the dot with $N$ electrons is stable, so that
addition/extraction energies,
$E^{\pm} \equiv E(N \pm 1) -E(N) \mp \mu$
where $\mu$ is the Fermi energy in the leads,
are positive. We are interested in the case where
$E^{\pm} \gg |\Delta|$, $\delta$ and also exceed the level broadening
$\Gamma^{i}=\pi\nu V_{i}^2$
and temperature $T$ (Coulomb blockade region).
In this case we can integrate out
the states with one or three extra electrons. This is equivalent to
Schrieffer-Wolff transformation which is used to obtain the conventional
Kondo model.\cite{classic2,classic3}
We obtain the following effective low-energy Hamiltonian
\begin{equation}
H_{\rm{eff}}=H_{\rm{leads}}+H_{\rm{dot}}+
H^{S=1}+H^{S=1 \leftrightarrow 0}+H_{\rm{eff}}^{\prime}.
\end{equation}
The Hamiltonian of the dot $H_{\rm{dot}}$ reads
\begin{equation}
H_{\rm{dot}}=\sum_{S,M} E_{SM} f_{SM}^{\dagger}f_{SM},
\end{equation}
using pseudo-fermion operators $f_{SM}^{\dagger}$
($f_{SM}$) which
create (annihilate) the state $|SM \rangle$. The condition of
\begin{equation}
\sum_{SM} f_{SM}^{\dagger} f_{SM} =1
\label{eq:constr0}
\end{equation}
should be fulfilled.
The third term $H^{S=1}$ represents the spin flip processes among
three components of the spin-triplet state. This resembles
a conventional Kondo Hamiltonian for $S=1$ in terms of the
spin operator $\hat{S}$
\begin{eqnarray}
H^{S=1} & = & \sum_{k k'} \sum_{i=1,2} J^{(i)}
\left[ \hat{S}_{+} c_{k' \downarrow}^{(i) \dagger} c_{k \uparrow}^{(i)}
+\hat{S}_{-} c_{k' \uparrow}^{(i) \dagger} c_{k \downarrow}^{(i)}
+\hat{S}_{z} (c_{k' \uparrow}^{(i) \dagger} c_{k \uparrow}^{(i)}
             -c_{k' \downarrow}^{(i) \dagger} c_{k \downarrow}^{(i)})
           \right] \nonumber \\
        & = & \sum_{k k'} \sum_{i=1,2} J^{(i)}
\Bigl[ \sqrt{2}(f_{11}^{\dagger}f_{10}+f_{10}^{\dagger}f_{1 -1})
                  c_{k' \downarrow}^{(i) \dagger} c_{k \uparrow}^{(i)}
+\sqrt{2}(f_{10}^{\dagger}f_{11}+f_{1 -1}^{\dagger}f_{10})
                  c_{k' \uparrow}^{(i) \dagger} c_{k \downarrow}^{(i)}
            \nonumber \\
& & +(f_{11}^{\dagger}f_{11}-f_{1 -1}^{\dagger}f_{1 -1})
                  (c_{k' \uparrow}^{(i) \dagger} c_{k \uparrow}^{(i)}
               -c_{k' \downarrow}^{(i) \dagger} c_{k \downarrow}^{(i)})
\Bigr].
\label{eq:H1}
\end{eqnarray}
The exchange coupling $J^{(i)}$ is accompanied by the scattering of
conduction electrons of channel $i$.
The fourth term $H^{S=1 \leftrightarrow 0}$ in $H_{\rm{eff}}$
describes the conversion between
the spin-triplet and singlet states accompanied by the interchannel
scattering of conduction electrons
\begin{eqnarray}
H^{S=1 \leftrightarrow 0} =
\sum_{k k'} \tilde{J}
\Bigl[\sqrt{2}(f_{11}^{\dagger}f_{00}-f_{00}^{\dagger}f_{1 -1})
                 c_{k' \downarrow}^{(1) \dagger} c_{k \uparrow}^{(2)}
+\sqrt{2}(f_{00}^{\dagger}f_{11}-f_{1 -1}^{\dagger}f_{00})
                 c_{k' \uparrow}^{(1) \dagger} c_{k \downarrow}^{(2)}
\nonumber \\
-(f_{10}^{\dagger}f_{00}+f_{00}^{\dagger}f_{10})
                 (c_{k' \uparrow}^{(1) \dagger} c_{k \uparrow}^{(2)}
                 -c_{k' \downarrow}^{(1) \dagger} c_{k \downarrow}^{(2)})
+ (1 \leftrightarrow 2)  \Bigr].
\label{eq:H2}
\end{eqnarray}
The coupling constants are given by
\begin{eqnarray}
J^{(i)} & = & \frac{V_{i}^2}{2E_{\rm c}}, \\
\tilde{J} & = & \frac{V_{1} V_{2}}{2E_{\rm c}},
\end{eqnarray}
where $1/E_{\rm c}=1/E^+ +1/E^-$. Note that $\tilde{J}^2=J^{(1)}J^{(2)}$.
The last term $H_{\rm{eff}}^{\prime}$ represents the scattering
processes of conduction electrons without any change of the
dot state and is not relevant for the current discussion.
The spin-flip processes included in our model are shown in Fig.\ 1(b).

\section{Scaling Calculations}

In this section we calculate the Kondo temperature $T_{\rm K}$
using the poor man's scaling technique.\cite{Anderson,multiK,multiK2}
By changing the energy scale (bandwidth of the conduction electrons)
from $D$ to $D-|d D|$, we obtain the scaling equations
using the second-order perturbation calculations with
respect to the exchange couplings, $J^{(1)}$, $J^{(2)}$, and $\tilde{J}$.
With decreasing $D$, the exchange couplings are renormalized.
The Kondo temperature is determined as the energy scale
at which the exchange couplings become so large that the perturbation
breaks down.

\subsection{In the absence of Zeeman effect}

When the Zeeman splitting is small and irrelevant, $E_{\rm Z} \ll T_{\rm K}$,
we obtain a closed form of the scaling equations for
$J^{(1)}$, $J^{(2)}$, and $\tilde{J}$ in two limits.\cite{me}
(i) When the energy scale $D$ is much larger than the energy
difference $|\Delta|$, $H_{\rm{dot}}$ can be safely disregarded
in $H_{\rm{eff}}$. The scaling equations can be written as
\begin{equation}
\frac{d}{d\ln D}
           \left( \begin{array}{cc}
           J^{(1)} & \tilde{J} \\
           \tilde{J} & J^{(2)} \end{array} \right)
=-2 \nu
           \left( \begin{array}{cc}
           J^{(1)} & \tilde{J} \\
           \tilde{J} & J^{(2)} \end{array} \right)^2.
\label{eq:scalA}
\end{equation}
(ii) For $D \ll \Delta$, the ground state of the dot is a spin triplet
and the singlet state can be disregarded. Then
$J^{(1)}$ and $J^{(2)}$ evolve independently 
\begin{equation}
\frac{d}{d\ln D} J^{(i)} = -2\nu J^{(i) 2},
\label{eq:scalB}
\end{equation}
whereas $\tilde J$ does not change.

In the case of $|\Delta| \ll T_{\rm K}$, the scaling equations
(\ref{eq:scalA}) remain valid till the scaling ends.
The matrix in Eq.\ (\ref{eq:scalA}) has eigenvalues of
\begin{eqnarray}
J_{\pm} & = & (J^{(1)}+J^{(2)})/2 \pm
\sqrt{(J^{(1)}-J^{(2)})^2/4+\tilde{J}^2}  \nonumber \\
 & = & J^{(1)}+J^{(2)}, 0.
\end{eqnarray}
The larger one, $J_{+}$, diverges upon decreasing the bandwidth $D$ and
determines $T_{\rm K}$
\begin{eqnarray}
T_{\rm K}(0) & = & D_0 \exp [-1/2\nu J_+]  \nonumber \\
 & = & D_0 \exp [-1/2\nu (J^{(1)}+J^{(2)})].
\end{eqnarray}
Here $D_0$ is the initial bandwidth, which is given by
$\sqrt{E^+ E^-}$.\cite{Haldane}

When $\Delta > D_0$, the scaling equations (\ref{eq:scalB}) work
in the whole scaling region. This yields
\begin{equation}
T_{\rm K}(\infty) =D_0 \exp [-1/2\nu J^{(1)}]
\end{equation}
for $J^{(1)} \ge J^{(2)}$.
This is the Kondo temperature for spin-triplet localized
spins.\cite{Okada}

In the intermediate region of $T_{\rm K}(0) \ll \Delta \ll D_0$, the
exchange couplings develop by Eq.\ (\ref{eq:scalA}) for 
$D \gg \Delta$. Around $D = \Delta$, $\tilde J$ saturates
while $J^{(1)}$ and $J^{(2)}$ continue to grow with decreasing
$D$, following Eq.\ (\ref{eq:scalB}) for $D \ll \Delta$.
We match the solutions of these scaling equations
at $D \simeq \Delta$ and obtain a power law of $T_{\rm K}(\Delta)$
\begin{equation}
T_{\rm K}(\Delta)=T_{\rm K}(0)\cdot 
\left( T_{\rm K}(0)/\Delta \right)^{\tan^2\theta},
\label{eq:TK}
\end{equation}
with
\begin{eqnarray}
\tan\theta & = &
\tilde{J}/[\sqrt{ (J^{(1)}-J^{(2)})^2/4+\tilde{J}^2}+
(J^{(1)}-J^{(2)})/2] \nonumber \\
& = & \sqrt{J^{(2)}/J^{(1)}}
\end{eqnarray}
for $J^{(1)} \ge J^{(2)}$.
Here $(\cos \theta, \sin\theta)^{T}$ is the eigenfunction of
the matrix in Eq.\ (\ref{eq:scalA}) corresponding to $J_+$.
$\theta\sim 0$ for $J^{(1)} \gg J^{(2)}$ and $\theta=\pi/4$
for $J^{(1)}=J^{(2)}$. In general, $0<\theta \le \pi/4$ and thus
$0 <\tan^2\theta \le 1$.

Finally, for $\Delta<0$, all the coupling constants saturate
and no Kondo effect is expected, provided $|\Delta| \gg T_{\rm K}(0)$.
Thus $T_{\rm K}$ quickly decreases to zero at $\Delta \sim -T_{\rm K}(0)$.
The Kondo temperature as a function of $\Delta$ is schematically shown
in Fig.\ 2(a).

\subsection{Case of $E_{\rm Z} =-\Delta$}

When $E_{\rm Z} =-\Delta$, the energies of states $|0 0 \rangle$ and
$|1 1 \rangle$ are degenerate. Then the Kondo effect is expected
even when $|\Delta| \gg T_{\rm K}(0)$.\cite{Pustilnik,Giuliano}
In this subsection
we evaluate $T_{\rm K}$ in this special case of $E_{\rm Z} =-\Delta$
by the poor man's scaling method.

(i) For the energy scale of $D \gg |\Delta|=E_{\rm Z}$,
$H_{\rm{dot}}$ can be disregarded in $H_{\rm eff}$.
The exchange couplings, $J^{(1)}$, $J^{(2)}$, and $\tilde{J}$, evolve
following Eq.\ (\ref{eq:scalA}).
(ii) In another limit of $D \ll |\Delta|=E_{\rm Z}$,
only the states $|0 0 \rangle$ and $|1 1 \rangle$ are relevant.
In $H_{\rm{eff}}$,
\begin{eqnarray}
H_{\rm{eff}}^{|0 0 \rangle,|1 1 \rangle}=
\sum_{k k'} \sum_{i=1,2} 
\Bigl[ \frac{1}{2} J_{\rm{s}}^{(i)}
(f_{11}^{\dagger}f_{11}-f_{00}^{\dagger}f_{00})
(c_{k' \uparrow}^{(i) \dagger} c_{k \uparrow}^{(i)}
 -c_{k' \downarrow}^{(i) \dagger} c_{k \downarrow}^{(i)})
+ \frac{1}{2} J_{\rm{c}}^{(i)}
(f_{11}^{\dagger}f_{11}+f_{00}^{\dagger}f_{00})
(c_{k' \uparrow}^{(i) \dagger} c_{k \uparrow}^{(i)}
 -c_{k' \downarrow}^{(i) \dagger} c_{k \downarrow}^{(i)}) \Bigr]
\nonumber \\
+\sum_{k k'} \sqrt{2}\tilde{J} \Bigl[
     f_{11}^{\dagger}f_{00}
      c_{k' \downarrow}^{(1) \dagger} c_{k \uparrow}^{(2)}
    +f_{00}^{\dagger}f_{11}
      c_{k' \uparrow}^{(1) \dagger} c_{k \downarrow}^{(2)}
+(1 \leftrightarrow 2) \Bigr].
\end{eqnarray}
$J_{\rm{s}}^{(i)}=J_{\rm{c}}^{(i)}=J^{(i)}$ initially.
The scaling procedure yields
\begin{equation}
\left\{
\begin{array}{l}
\frac{d}{d\ln D} J_{\rm{s}}^{(i)}=-4 \nu \tilde{J}^2 \\
\frac{d}{d\ln D} \tilde{J}=-\nu (J_{\rm{s}}^{(1)}+J_{\rm{s}}^{(2)})
\tilde{J},
\end{array}\right.
\label{eq:scalC}
\end{equation}
and $J_{\rm{c}}^{(i)}$ do not change. These scaling equations
are nearly equivalent to those of the anisotropic Kondo model with
$S=1/2$,\cite{Anderson} as pointed out in Refs.\
\onlinecite{Pustilnik,Giuliano}.

When $|\Delta|=E_{\rm Z}>D_0$, the scaling equations
(\ref{eq:scalC}) remain valid in the whole scaling region.
This yields the Kondo temperature
\begin{equation}
T_{\rm K}(\infty) =D_0 \exp [-A(\theta)/2\nu (J^{(1)}+J^{(2)})]
\end{equation}
with
\begin{equation}
A(\theta)=
\left\{
\begin{array}{ll}
\frac{1}{\lambda}\ln\left( \frac{1+\lambda}{1-\lambda} \right)
& \quad (0< \theta \le \pi/8)
\\
\frac{2}{\lambda} \tan^{-1} \lambda
& \quad (\pi/8 < \theta \le \pi/4),
\end{array}\right.
\end{equation}
where $\displaystyle \lambda=\sqrt{|\cos 4\theta|}$.
$A(\theta)$ decreases monotonically with increasing $\theta$.
$A(\theta) \rightarrow \infty$ as $\theta \rightarrow 0$.
$A(\pi/8)=2$ and $A(\pi/4)=\pi/2$. When $J^{(1)}+J^{(2)}$ is
fixed, $T_{\rm K}(\infty)$ is the largest at $J^{(1)}=J^{(2)}$
($\theta=\pi/4$) and becomes smaller with decreasing
$J^{(2)}/J^{(1)} (=\tan^2\theta)$.

In the intermediate region,
$T_{\rm K}(0) \ll |\Delta|=E_{\rm Z} \ll D_0$, we match
the solutions of Eqs.\ (\ref{eq:scalA}) and (\ref{eq:scalC})
at $D \simeq |\Delta|$. We obtain a power law
\begin{equation}
T_{\rm K}(\Delta)=T_{\rm K}(0)\cdot \left( T_{\rm K}(0)/|\Delta|
\right)^{A(\theta)-1}.
\label{eq:TK2}
\end{equation}
Figure 2(b) shows the behaviors of $T_{\rm K}(\Delta)$ in the case of
$E_{\rm Z}=-\Delta$.

\section{Mean Field Calculations}

\subsection{Kondo resonance for spin $S=1/2$}

To illustrate the mean field theory for the Kondo effect in quantum dots,
we begin with the usual case of $S=1/2$. We assume that
one level ($E_0$) in a quantum dot is occupied by an electron
with spin either up or down ($\sigma=\uparrow,\downarrow$).
The effective low-energy Hamiltonian is
\begin{equation}
H=\sum_{k \sigma} \varepsilon_{k} c_{k\sigma}^{\dagger} c_{k\sigma}
+\sum_{\sigma} E_{\sigma} f_{\sigma}^{\dagger}f_{\sigma}
+J \sum_{k k'} \sum_{\sigma,\sigma'} f_{\sigma}^{\dagger}f_{\sigma'}
                                 c_{k'\sigma'}^{\dagger} c_{k\sigma},
\label{eq:Ham0}
\end{equation}
with the constraint of
\begin{equation}
f_{\uparrow}^{\dagger}f_{\uparrow}
+f_{\downarrow}^{\dagger}f_{\downarrow}=1.
\label{eq:constraint}
\end{equation}
For electrons in leads $L$, $R$, we have performed a unitary
transformation of
$c_{k \sigma}=(V_{L}^*c_{L,k \sigma}+V_{R}^*c_{R,k \sigma})/
\sqrt{|V_{L}|^2+|V_{R}|^2}$ where $V_{\alpha}$ is the tunneling
coupling to lead $\alpha$.\cite{Glazman} The last term in
Eq.\ (\ref{eq:Ham0}) represents the exchange coupling between $S=1/2$
in the dot and conduction electrons (Appendix A).

In the mean field theory, we introduce the order parameter
\begin{equation}
\langle \Xi \rangle = \frac{1}{\sqrt{2}}
\sum_{k}
(\langle f_{\uparrow}^{\dagger} c_{k\uparrow} \rangle +
\langle f_{\downarrow}^{\dagger} c_{k\downarrow}  \rangle)
\end{equation}
to describe the spin couplings between the dot states and conduction
electrons. The mean field Hamiltonian reads
\begin{equation}
H_{\rm MF}=
\sum_{k \sigma} \varepsilon_{k} c_{k\sigma}^{\dagger} c_{k\sigma}
+\sum_{\sigma} E_{\sigma} f_{\sigma}^{\dagger}f_{\sigma}
-\sum_{k,\sigma} (\sqrt{2}J \langle \Xi \rangle 
c_{k \sigma}^{\dagger} f_{\sigma}+{\rm H.c.})
+2J |\langle \Xi \rangle|^2
+\lambda \left(\sum_{\sigma} f_{\sigma}^{\dagger}f_{\sigma}-1 \right).
\label{eq:Ham0MF}
\end{equation}
The constraint, Eq.\ (\ref{eq:constraint}), is taken into account by
the last term with a Lagrange multiplier $\lambda$.
By minimizing the expectation value of $H_{\rm MF}$,
$\langle \Xi \rangle$ is determined self-consistently (see Appendix A).

In the absence of the Zeeman effect,
$E_{\uparrow}=E_{\downarrow}=E_0$.
The mean field Hamiltonian, $H_{\rm MF}$, represents a resonant
tunneling through an ``energy level,'' $\tilde{E}_0=E_0+\lambda$,
with ``tunneling coupling,'' $\tilde{V}=-\sqrt{2}J \langle \Xi \rangle$.
$\tilde{V}$ provides a finite width of the resonance, $\tilde{\Delta}_0=
\pi \nu |\tilde{V}|^2$, with $\nu$ being the density of states in the leads.
The constraint, Eq.\ (\ref{eq:constraint}), requires that the
states for the pseudo-fermions are half-filled, that is,
$\tilde{E}_0=\mu$. Hence the Kondo resonant state appears just at the
Fermi level $\mu$,
as indicated in the inset (A) in Fig.\ 3(a).
The self-consistent calculations give us the
resonant width
\begin{equation}
\tilde{\Delta}_0=
\pi \nu \left| \sqrt{2}J \langle \Xi \rangle \right|^2=
D_0 \exp [-1/2 \nu J].
\label{eq:TK1}
\end{equation}
This is identical to the Kondo temperature $T_{\rm K}$.

In the presence of the Zeeman splitting,
$E_{\uparrow}=E_0 - E_{\rm Z}$ and $E_{\downarrow}=E_0 + E_{\rm Z}$.
Hence the resonant level is split for spin-up and down electrons,
$\tilde{E}_{\uparrow / \downarrow}=E_{\uparrow / \downarrow}+\lambda$.
The constraint, Eq.\ (\ref{eq:constraint}), yields $E_0+\lambda=\mu$
(see inset (B) in Fig.\ 3(a)).
The resonant width $\tilde{\Delta}$ is determined as
\begin{equation}
\tilde{\Delta}^2+E_{\rm Z}^2=\tilde{\Delta}_0^2,
\end{equation}
where $\tilde{\Delta}_0$ is given by Eq.\ (\ref{eq:TK1}).
The Kondo temperature is evaluated by this width,
$T_{\rm K}(E_{\rm Z})=\tilde{\Delta}$.
$T_{\rm K}$ decreases with increasing $E_{\rm Z}$ and
disappears at $E_{\rm Z}=T_{\rm K}(0)$, as shown in Fig.\ 3(a).

The conductance $G$ through the dot is expressed, using
$\Gamma_{\alpha}=\pi\nu |V_{\alpha}|^2$, as
\begin{equation}
G = \frac{2e^2}{h} \frac{4\Gamma_L\Gamma_R}{(\Gamma_L+\Gamma_R)^2}
\left[1-\left( \frac{E_{\rm Z}}{T_{\rm K}(0)} \right)^2 \right].
\label{eq:cond}
\end{equation}
This is the conductance in the unitary limit for $E_{\rm Z}=0$.
Figure 3(b) presents the $E_{\rm Z}$ dependence of the conductance.
With increasing $E_{\rm Z}$, the splitting between
the resonant levels for spin up and
down becomes larger. In consequence the amplitude of the Kondo
resonance decreases at $\mu$, which reduces the conductance.

\subsection{Kondo resonance in the present model}

Now we apply the mean field theory to our model which has
the spin-triplet and singlet states in a quantum dot. The spin states
of the coupling to a conduction electron are
$(S=1)\otimes(S=1/2)=(S=3/2)\oplus(S=1/2)$ for the former, and
$(S=0)\otimes(S=1/2)=(S=1/2)$ for the latter (Appendix B).
To represent the competition between the triplet and single states,
therefore, the order parameter should be a spinor of $S=1/2$.
It is $\langle \vec{\Xi} \rangle$ where
\begin{equation}
\vec{\Xi}  =  \sum_k \left( \begin{array}{c}
        \cos\varphi\left( \sqrt{2}f_{11}^{\dagger}c_{k \uparrow}^{(1)} + 
              f_{10}^{\dagger}c_{k \downarrow}^{(1)}  \right)/\sqrt{3} +
        \sin\varphi f_{00}^{\dagger} c_{k \downarrow}^{(2)}  \\
     \cos\varphi\left( \sqrt{2}f_{1-1}^{\dagger}c_{k \downarrow}^{(1)} + 
                f_{10}^{\dagger}c_{k \uparrow}^{(1)}  \right)/\sqrt{3} -
        \sin\varphi f_{00}^{\dagger} c_{k \uparrow}^{(2)}
        \end{array} \right)
\label{eq:orderp}
\end{equation}
for $J^{(1)}>J^{(2)}$. A mode of the largest coupling is taken into
account in this approximation. The Hamiltonian reads
\begin{equation}
H_{\rm MF}=H_{\rm lead}+H_{\rm dot}
-J_{\rm MF} \left[ \langle \vec{\Xi}^{\dagger} \rangle \vec{\Xi} + 
                \vec{\Xi}^{\dagger} \langle \vec{\Xi} \rangle -
                |\langle \vec{\Xi} \rangle|^2 \right]
+\lambda \left( \sum_{SM} f_{SM}^{\dagger}f_{SM}-1 \right),
\label{eq:HamMF}
\end{equation}
where
\begin{equation}
J_{\rm MF}=J^{(1)}+\sqrt{J^{(1) 2}+3\tilde{J}^2},
\label{eq:MFev}
\end{equation}
and
\begin{equation}
\tan\varphi=\sqrt{3}\tilde{J}/J_{\rm MF}.
\label{eq:MFef}
\end{equation}
The last term in $H_{\rm MF}$ considers the restriction of
Eq.\ (\ref{eq:constr0}).
The expectation value of $H_{\rm MF}$ is minimized with respect to
$|\vec{\Xi}|^2$.
The Kondo temperature can be estimated by
\begin{equation}
T_{\rm K}= \pi\nu |J_{\rm MF}\langle \vec{\Xi} \rangle|^2,
\end{equation}
using $\langle \vec{\Xi} \rangle$ determined by the self-consistent
calculations (Appendix B).

First let us consider the case in the absence of the Zeeman effect,
$E_{1 M}=E_{S=1}$ and $E_{00}=E_{S=1}+\Delta$.
The resonant level for the triplet state is threefold degenerate
at $\tilde{E}_{S=1}=E_{S=1}+\lambda$ whereas the resonant level for
the singlet state is at $\tilde{E}_{0}=E_{00}+\lambda$. These levels
are separated by the energy $\Delta$. The Lagrange multiplier
$\lambda$ is determined to fulfill Eq.\ (\ref{eq:constr0}).
Figure 4(a) shows the calculated results of $T_{\rm K}$ as a function of
$\Delta$. Both of $T_{\rm K}$ and $\Delta$ are in units of
$D_0 \exp (-1/\nu J_{\rm MF})$. We find that
(i) $T_{\rm K}(\Delta)$ reaches its maximum at $\Delta=0$, (ii) for
$\Delta \gg T_{\rm K}(0)$, $T_{\rm K}(\Delta)$ obeys a power law
\begin{equation}
T_{\rm K}(\Delta) \cdot \Delta^{\tan^2\varphi}=\rm{const}.,
\label{eq:power1}
\end{equation}
and (iii) for $\Delta <0$, $T_{\rm K}$ decreases rapidly with increasing
$|\Delta|$ and disappears at
$\Delta=\Delta_{\rm c} \sim -T_{\rm K}(0)$;
\begin{equation}
\Delta_{\rm c}=-D_0 \exp (-1/\nu J_{\rm MF}) (1+\tan^2\varphi)
(\tan^2\varphi)^{-\sin^2\varphi}.
\end{equation}
These features are in agreement with the results of the
scaling calculations.

The behaviors of $T_{\rm K}(\Delta)$ can be understood as follows.
The inset of Fig.\ 4(a) schematically shows the Kondo resonant states.
The resonance of the triplet state is denoted by solid lines whereas
that of the singlet state is by dotted lines.
(A) When $\Delta \gg T_{\rm K}(0)$, the triplet resonance appears around
$\mu$ whereas the singlet resonance is far above $\mu$.
(B) With a decrease in $\Delta$, the two resonant states
are more overlapped at $\mu$, which raises $T_{\rm K}$ gradually. This
results in a power law of $T_{\rm K}(\Delta)$, Eq.\ (\ref{eq:power1}).
The largest overlap yields the maximum of $T_{\rm K}$ at $\Delta=0$.
(C) When $\Delta<0$, the singlet and triplet resonances are located
below and above $\mu$, respectively, being sharper and farther from
each other with increasing $|\Delta|$.
Finally the Kondo resonance disappears at $\Delta=\Delta_{\rm c}$.

The conductance through the dot is given by
\begin{equation}
G / (e^2/h) = \left.
\frac{4\Gamma_L^{1}\Gamma_R^{1}}{(\Gamma_L^{1}+\Gamma_R^{1})^2}
\left( \frac{\tilde{\Delta}_{11}^2}{(\varepsilon-\tilde{E}_{11})^2+
\tilde{\Delta}_{11}^2}
+\frac{\tilde{\Delta}_{10}^2}{(\varepsilon-\tilde{E}_{10})^2+
\tilde{\Delta}_{10}^2}
\right)
+\frac{4\Gamma_L^{2}\Gamma_R^{2}}{(\Gamma_L^{2}+\Gamma_R^{2})^2}
\frac{\tilde{\Delta}_{00}^2}{(\varepsilon-\tilde{E}_{00})^2+
\tilde{\Delta}_{00}^2}
\right|_{\varepsilon=\mu},
\label{eq:cond2}
\end{equation}
where $\Gamma_{\alpha}^{i}=\pi\nu |V_{\alpha,i}|^2$. The
resonant widths are
$\tilde{\Delta}_{11}/\tilde{\Delta}_0=2\cos^2\varphi/3$,
$\tilde{\Delta}_{10}/\tilde{\Delta}_0=\cos^2\varphi/3$, and
$\tilde{\Delta}_{00}/\tilde{\Delta}_0=\sin^2\varphi$ with
$\tilde{\Delta}_0=
\pi \nu | J_{\rm MF} \langle \vec{\Xi} \rangle |^2$.
The conductance $G$ as a function of $\Delta$ is shown in Fig.\ 4(b),
in a symmetric case of 
$\Gamma_L^{i}=\Gamma_R^{i}$ ($i=1,2$).
$G=2e^2/h$ for $\Delta>0$ whereas $G$ goes to zero suddenly
for $\Delta<0$. Around $\Delta=0$, $G$ is larger than the
value in the unitary limit, $2e^2/h$,
which is attributable to nonuniversal
contribution from the multichannel nature of our model.\cite{me}

In the presence of the Zeeman splitting, $E_{1 M}=E_{S=1}-E_{\rm Z} M$,
the resonant level of the triplet state is split into three.
With increasing $E_{\rm Z}$, the Kondo effect is rapidly weaken
except in the region of $\Delta \sim -E_{\rm Z}$.
In Fig.\ 5(a), we show the Kondo temperature $T_{\rm K}$ in
$E_{\rm Z}$-$\Delta$ plane, in the case of $\varphi=0.15\pi$.
Figure 5(b) presents $T_{\rm K}$ as a function of $\Delta$ for several
values of $E_{\rm Z}$. When $E_{\rm Z}$ is large enough,
the Kondo effect takes place only when the resonant state of
$|1 1 \rangle$ is overlapped with that of $|0 0 \rangle$. Then
$T_{\rm K}$ is the largest at $\Delta=-E_{\rm Z}$ and decreases with
$\Delta$ being away from this value.
At $\Delta=-E_{\rm Z}$, $T_{\rm K}$ obeys a power law
\begin{equation}
T_{\rm K}(\Delta) \cdot |\Delta|^{1/(2+3\tan^2\varphi)}={\rm const}.,
\label{eq:power}
\end{equation}
which is indicated by a broken line in Fig.\ 5(b).
This is qualitatively in agreement with the calculated results by the
scaling method.

Figure 6 indicates the conductance $G$ in $E_{\rm Z}$-$\Delta$ plane,
when $\varphi=0.15\pi$ and $\Gamma_L^{i}=\Gamma_R^{i}$ ($i=1,2$).
$G$ takes the value of $2e^2/h$ around $E_{\rm Z}=0$ and $\Delta>0$,
and also along the line of $E_{\rm Z}=-\Delta$. ($G>2e^2/h$ in the
neighborhood of $E_{\rm Z}=\Delta=0$, as discussed above.)
For sufficiently large $E_{\rm Z}$, our model is nearly
equivalent to the anisotropic Kondo model with
$S=1/2$.\cite{Pustilnik,Giuliano} Hence $G=2e^2/h$ at $\Delta=-E_{\rm Z}$
and reduces to zero as $\Delta$ deviates from this value, in the same
way as in Fig.\ 3(b) for the case of $S=1/2$.

\section{Conclusions and Discussion}

The Kondo effect in quantum dots with an even number of electrons
has been investigated theoretically. The Kondo temperature $T_{\rm K}$
has been calculated as a function of the energy difference
$\Delta=E_{00}-E_{S=1}$ and the Zeeman splitting $E_{\rm Z}$,
using the poor man's scaling method and mean field theory.
The scaling calculations have indicated that
the competition between the spin-triplet and singlet states
significantly enhances the Kondo effect.
When the Zeeman effect is irrelevant, $E_{\rm Z} \ll T_{\rm K}$,
$T_{\rm K}$ is maximal around $\Delta=0$ and decreases with
$\Delta$ obeying a power law. In a case of
$-\Delta = E_{\rm Z}$, the Kondo effect takes place from the
degeneracy between two states, $|00 \rangle$ and $|11 \rangle$.
Even in this case, the contribution from the other states of
higher energy, $|10 \rangle$ and $|1-1 \rangle$, plays an
important role in the enhancement of $T_{\rm K}$.
As a result, $T_{\rm K}$ is maximal around
$E_{\rm Z}=0$ and depends on $E_{\rm Z}$ by a power law again.

The mean field theory yields a clear cut view for the
Kondo effect in quantum dots. Considering the spin couplings
between the dot states and conduction electrons as a
mean field, $\langle f_{SM}^{\dagger}c_{k,\sigma}^{(i)}\rangle$,
we find that the resonant states are
created around the Fermi level $\mu$.
The resonant width is given by the Kondo temperature $T_{\rm K}$.
The unitary limit of the conductance, $G\sim 2e^2/h$, can be
easily understood in terms of the tunneling through these resonant
states. In our model,
the overlap between the resonant states of $S=1$ and $S=0$ in the
dot enhances the Kondo effect.
The mean field calculations have led to a power law
dependence of $T_{\rm K}$ on $\Delta$ and on $E_{\rm Z}$, in accordance
with the scaling calculations.

The mean field theory is not quantitatively accurate for the evaluation
of $T_{\rm K}$.\cite{com0} (In the case of $S=1/2$, the exact value of
$T_{\rm K}$ is obtained accidentally.) In our model,
the scaling calculations indicate that all the
exchange couplings, $J^{(1)}$, $J^{(2)}$, and $\tilde{J}$, are
renormalized altogether following Eq.\ (\ref{eq:scalA}) when
$|\Delta|$ and $E_{\rm Z}$ are much smaller than the energy scale $D$.
In consequence two channels in the leads are coupled effectively for
an increase in $T_{\rm K}$.
In the mean field calculations, the interchannel couplings are taken
into account in Eq.\ (\ref{eq:MFev}) only partly. In fact, conduction
electrons of channel $1$ and $2$ independently take part in the
conductance, Eq.\ (\ref{eq:cond2}). By the perturbation calculations
with respect to the exchange couplings, we find that mixing terms
between the channels appear in the logarithmic corrections to the
conductance.\cite{me} We could improve the mean field calculations
by adopting another form of the order parameter than
Eq.\ (\ref{eq:orderp}).

Our calculated results explain the experimental findings by
Sasaki {\it et al}.:\cite{Sasaki} The Kondo effect is largely
enhanced around $\Delta=0$ when the Zeeman effect is irrelevant.
The behavior of $T_{\rm K}$ in the presence of the Zeeman effect may be
observed experimentally under higher magnetic fields. In experiments
the value of the Zeeman splitting can be controlled by applying a
magnetic field parallel to the quantum dot.
More generally, it is possible to control several parameters in
semiconductor quantum dots and to realize new situations which
cannot be reached in traditional solid state context.
The quantum dot systems, therefore, have the potential of tools
to explore the Kondo physics further beyond the present theory.

%---------------------------------------------
\section*{ACKNOWLEDGMENTS}

The authors are indebted to L.\ P.\ Kouwenhoven, S.\ De Franceschi,
J.\ M.\ Elzerman, K.\ Maijala, S.\ Sasaki,
W.\ G.\ van der Wiel, Y.\ Tokura, L.\ I.\ Glazman, M.\ Pustilnik,
and G.\ E.\ W.\ Bauer for
valuable discussions.
The authors acknowledge financial support from the
``Netherlandse Organisatie voor
Wetenschappelijk Onderzoek'' (NWO). M.\ E.\ is also grateful for financial
support from the
Japan Society for the Promotion of Science for his stay at Delft University
of Technology.

%--------------------------------------------- Appendix
\appendix

\section{Mean field calculations for $S=1/2$}

The original Hamiltonian for a quantum dot with one energy level
reads
\begin{equation}
H=\sum_{\alpha=L,R}\sum_{k \sigma}
\varepsilon_{k} c_{\alpha, k\sigma}^{\dagger}
c_{\alpha, k\sigma}+
\sum_{\alpha=L,R} \sum_{k \sigma} (V_{\alpha}c_{\alpha,
k\sigma}^{\dagger} d_{\sigma} + {\rm H.c.})+H_{\rm dot}
\end{equation}
with
\begin{equation}
H_{\rm dot}=\sum_{\sigma} E_0 d_{\sigma}^{\dagger} d_{\sigma}
+U d_{\uparrow}^{\dagger} d_{\uparrow} d_{\downarrow}^{\dagger} d_{\downarrow}.
\label{eq:H0dot}
\end{equation}
For the state of one electron in the dot, the addition and extraction
energies are given by $E^+=E_0+U-\mu$ and $E^-=\mu-E_0$, respectively.
The parameters, $E_0$ and $U$, in Eq.\ (\ref{eq:H0dot})
should be determined to fit these energies to experimental data.
For conduction electrons in leads $L$, $R$, we perform a unitary
transformation, $c_{k \sigma}=(V_{L}^*c_{L,k \sigma}+
V_{R}^*c_{R,k \sigma})/V$,
$\bar{c}_{k \sigma}=(-V_{R,i}c_{L,k \sigma}+V_{L,i}c_{R,k \sigma})/V$,
with $V=\sqrt{|V_{L}|^2+|V_{R}|^2}$, along the lines of
Ref.\ \onlinecite{Glazman}.
We disregard the modes $\bar{c}_{k \sigma}$ which are uncoupled to
the quantum dot.

We consider the Coulomb blockade region for one electron, where
both $E^+$ and $E^-$
are much larger than the level broadening $\Gamma=\pi\nu V^2$
($\nu$ being the density of states in the leads) and temperature.
Integrating out the dot states with zero or two electrons by
the Schrieffer-Wolff transformation,\cite{classic2,classic3}
we obtain the effective low-energy Hamiltonian
\begin{equation}
H=\sum_{k \sigma} \varepsilon_{k} c_{k\sigma}^{\dagger} c_{k\sigma}
+\sum_{\sigma} E_{\sigma} f_{\sigma}^{\dagger}f_{\sigma}
+J \sum_{k k'} 
\left[ \hat{S}_{+} c_{k' \downarrow}^{\dagger} c_{k \uparrow}
+\hat{S}_{-} c_{k' \uparrow}^{\dagger} c_{k \downarrow}
+\hat{S}_{z} (c_{k' \uparrow}^{\dagger} c_{k \uparrow}
             -c_{k' \downarrow}^{\dagger} c_{k \downarrow})
           \right]
\label{eq:Ham00}
\end{equation}
under a constraint of Eq.\ (\ref{eq:constraint}). In the second term
we have included the Zeeman effect, $E_{\uparrow,\downarrow}=
E_0 \pm E_{\rm Z}$.
The third term represents the exchange coupling between the dot spin
and conduction electrons with $J=V^2/E_{\rm c}$ where $1/E_{\rm c}=1/E^+ +1/E^-$.
By expressing the spin operator $\hat{S}$ as
$\hat{S}_{+}=f_{\uparrow}^{\dagger} f_{\downarrow}$,
$\hat{S}_{-}=f_{\downarrow}^{\dagger} f_{\uparrow}$,
$\hat{S}_{z}=(f_{\uparrow}^{\dagger} f_{\uparrow}-
              f_{\downarrow}^{\dagger} f_{\downarrow})/2$,
one finds that Eq.\ (\ref{eq:Ham00}) is identical to Eq.\
(\ref{eq:Ham0}).

The mean field Hamiltonian, Eq.\ (\ref{eq:Ham0MF}), includes
``energy levels'' for pseudo-fermions,
$\tilde{E}_{\sigma}=E_{\sigma}+\lambda$, which are coupled to the leads
with ``tunneling amplitude,''
$\tilde{V}=-\sqrt{2}J \langle \Xi \rangle$.
The Green function for the pseudo-fermions is
\begin{equation}
G_{\sigma} (\varepsilon)=\frac{1}{\varepsilon-\tilde{E}_{\sigma}+{\rm i}
\tilde{\Delta}},
\label{eq:GreenF}
\end{equation}
where $\tilde{\Delta}=\pi \nu |\tilde{V}|^2$. This represents the
resonant tunneling with the resonant width $\tilde{\Delta}$.

The expectation value of the Hamiltonian, Eq.\ (\ref{eq:Ham0MF}), is
written as
\begin{equation}
E_{\rm MF}=\sum_{\sigma} \left[
-\frac{\tilde{\Delta}}{\pi}+\frac{\tilde{E}_{\sigma}}{\pi}
\tan^{-1} \frac{\tilde{\Delta}}{\tilde{E}_{\sigma}}+
\frac{\tilde{\Delta}}{2\pi}
\ln \frac{\tilde{E}_{\sigma}^2+\tilde{\Delta}^2}{D_0^2}
\right] - \lambda + \frac{\tilde{\Delta}}{\pi\nu J},
\end{equation}
where $D_0$ is the bandwidth in the leads.\cite{classic2}
We set $\mu=0$ in this appendix.
The constraint of Eq.\ (\ref{eq:constraint}) is equivalent to the
condition
\begin{equation}
\frac{\partial E_{\rm MF}}{\partial \lambda}=
\frac{1}{\pi} \sum_{\sigma}
\tan^{-1} \frac{\tilde{\Delta}}{\tilde{E}_{\sigma}}-1=0.
\end{equation}
This yields $E_0+\lambda=0$. The minimization of $E_{\rm MF}$ with
respect to $\tilde{\Delta}$ (or $|\langle \Xi \rangle|^2$) determines
$\tilde{\Delta}$
\begin{equation}
\frac{\partial E_{\rm MF}}{\partial \tilde{\Delta}}=
\frac{1}{2\pi} \sum_{\sigma}
\ln \frac{\tilde{E}_{\sigma}^2+\tilde{\Delta}^2}{D_0^2}
+\frac{1}{\pi\nu J}=0.
\label{eq:MFeq}
\end{equation}
For $E_{\rm Z}=0$, we find
\begin{equation}
\tilde{\Delta}=D_0 \exp [-1/2 \nu J] \equiv \tilde{\Delta}_0.
\end{equation}
This is equal to the Kondo temperature, $T_{\rm K}$.
For $E_{\rm Z} \ne 0$, Eq.\ (\ref{eq:MFeq}) yields
\begin{equation}
\tilde{\Delta}^2+E_{\rm Z}^2=\tilde{\Delta}_0^2.
\end{equation}

Using the T-matrix, $\hat{T}$, the conductance through the dot, $G$,
is given by
\begin{eqnarray}
G & = & \frac{e^2}{h} (2\pi\nu)^2 \sum_{\sigma} \left.
|\langle R, k' \sigma | \hat{T} | L, k \sigma \rangle |^2
\right|_{\varepsilon_k=\varepsilon_{k'}=\mu} \nonumber \\
 & = & \frac{e^2}{h} (2\pi\nu)^2
 \frac{|V_L|^2 |V_R|^2}{(|V_L|^2+|V_R|^2)^2} \sum_{\sigma} \left.
|\langle \psi_{k' \sigma} | \hat{T} | \psi_{k \sigma} \rangle |^2
\right|_{\varepsilon_k=\varepsilon_{k'}=\mu} \nonumber \\
 & = & \frac{e^2}{h} 
 \frac{4 \Gamma_L \Gamma_R}{(\Gamma_L+\Gamma_R)^2}
 \sum_{\sigma} \left.
 \frac{\tilde{\Delta}^2}
 {(\varepsilon-\tilde{E}_{\sigma})^2+\tilde{\Delta}^2}
 \right|_{\varepsilon=\mu}
\label{eq:cond0}
\end{eqnarray}
where $\Gamma_{\alpha}=\pi\nu |V_{\alpha}|^2$.
This yields Eq.\ (\ref{eq:cond}) in the text.
On the second line in Eq.\ (\ref{eq:cond0}),
$| \psi_{k \sigma} \rangle =c_{k \sigma}^{\dagger} |0 \rangle =
(V_{L} |L, k \sigma \rangle+
V_{R} |R, k \sigma \rangle)/V$, and the T-matrix is
evaluated in terms of the Green function, Eq.\ (\ref{eq:GreenF}),
$|\tilde{V}|^2 G_{\sigma} (\varepsilon=\varepsilon_k)$.

\section{Mean field calculations in the present model}

For the spin states of the coupling between the spin triplet
$S=1$ in the dot and a conduction electron, we introduce spinors of
$S=1/2$ and $3/2$. Using the Clebsch-Gordan coefficients, they
are given by
\begin{eqnarray}
\vec{\Omega}_{1/2}^{(i)} & = & \sum_k \left( \begin{array}{c}
 (\sqrt{2}f_{11}^{\dagger}c_{k \uparrow}^{(i)} + 
        f_{10}^{\dagger}c_{k \downarrow}^{(i)})/\sqrt{3} \\
 (\sqrt{2}f_{1-1}^{\dagger}c_{k \downarrow}^{(i)} + 
        f_{10}^{\dagger}c_{k \uparrow}^{(i)})/\sqrt{3}
        \end{array} \right),
\\
\vec{\Omega}_{3/2}^{(i)} & = & \sum_k \left( \begin{array}{c}
          f_{11}^{\dagger}c_{k \downarrow}^{(i)} \\
 (-f_{11}^{\dagger}c_{k \uparrow}^{(i)} + \sqrt{2}
        f_{10}^{\dagger}c_{k \downarrow}^{(i)})/\sqrt{3} \\
 (f_{1-1}^{\dagger}c_{k \downarrow}^{(i)} - \sqrt{2}
        f_{10}^{\dagger}c_{k \uparrow}^{(i)})/\sqrt{3} \\
         -f_{1-1}^{\dagger}c_{k \uparrow}^{(i)}
        \end{array} \right).
\end{eqnarray}
The exchange couplings between the triplet state and
conduction electrons, Eq.\ (\ref{eq:H1}), can be rewritten as
\begin{equation}
H^{S=1}=\sum_{i=1,2} J^{(i)} \left[
-2\vec{\Omega}_{1/2}^{(i) \dagger} \vec{\Omega}_{1/2}^{(i)}
+\vec{\Omega}_{3/2}^{(i) \dagger} \vec{\Omega}_{3/2}^{(i)}
\right].
\end{equation}
In the same way we define the spinors of $S=1/2$ to represent the spin
couplings between the singlet state $S=0$ and a
conduction electron
\begin{equation}
\vec{\Psi}^{(i)}  =  \sum_k \left( \begin{array}{c}
f_{00}^{\dagger}c_{k \downarrow}^{(\bar{i})} \\
-f_{00}^{\dagger}c_{k \uparrow}^{(\bar{i})}
\end{array} \right),
\end{equation}
where $\bar{i}=2$ and $1$ for $i=1$ and $2$, respectively.
The conversion between the triplet and singlet states, Eq.\ (\ref{eq:H2}),
is rewritten as
\begin{equation}
H^{S=1 \leftrightarrow 0}=
-\sqrt{3} \tilde{J} \sum_{i=1,2}  \left[
\vec{\Psi}^{(i) \dagger} \vec{\Omega}_{1/2}^{(i)}
+\rm{H.c.} \right].
\end{equation}
In $H^{S=1}+H^{S=1 \leftrightarrow 0}$,
a mode of the largest coupling with $S=1/2$ is given by
\begin{equation}
\vec{\Xi} = \cos\varphi \vec{\Omega}_{1/2}^{(1)} +
        \sin\varphi \vec{\Psi}^{(1)}
\end{equation}
for $J^{(1)} \ge J^{(2)}$, which is Eq.\ (\ref{eq:orderp}) in the text.
The corresponding eigenvalue is given by Eq.\ (\ref{eq:MFev}) and
$\varphi$ is determined as in Eq.\ (\ref{eq:MFef}).

The mean field Hamiltonian, Eq.\ (\ref{eq:HamMF}), represents the
resonant tunneling through the energy levels for the pseudo-fermions,
$\tilde{E}_{SM}=E_{SM}+\lambda$.
The expectation value of Eq.\ (\ref{eq:HamMF}), $E_{\rm MF}$,
is evaluated in the same way as in Appendix A. 
$\partial E_{\rm MF}/\partial \lambda=0$ yields
\begin{equation}
\tan^{-1} \frac{\tilde{\Delta}_{11}}{\tilde{E}_{11}}+
\tan^{-1} \frac{\tilde{\Delta}_{10}}{\tilde{E}_{10}}+
\tan^{-1} \frac{\tilde{\Delta}_{00}}{\tilde{E}_{00}}
=\pi,
\label{eq:MFeq1}
\end{equation}
where the resonant widths are
$\tilde{\Delta}_{11}/\tilde{\Delta}_0=2\cos^2\varphi/3$,
$\tilde{\Delta}_{10}/\tilde{\Delta}_0=\cos^2\varphi/3$, and
$\tilde{\Delta}_{00}/\tilde{\Delta}_0=\sin^2\varphi$ with
$\tilde{\Delta}_0=\pi \nu | J_{\rm MF} \langle \vec{\Xi} \rangle |^2$.
We set $\mu=0$ here.
Minimizing $E_{\rm MF}$ with respect to $\tilde{\Delta}_0$, we obtain
\begin{equation}
\frac{2}{3}\cos^2\varphi
\ln \frac{\tilde{E}_{11}^2+\tilde{\Delta}_{11}^2}{D_0^2}
+\frac{1}{3}\cos^2\varphi
\ln \frac{\tilde{E}_{10}^2+\tilde{\Delta}_{10}^2}{D_0^2}
+\sin^2\varphi
\ln \frac{\tilde{E}_{00}^2+\tilde{\Delta}_{00}^2}{D_0^2}
+\frac{2}{\nu J}=0.
\label{eq:MFeq2}
\end{equation}
Equations (\ref{eq:MFeq1}) and (\ref{eq:MFeq2}) determine
$\lambda$ and $\tilde{\Delta}_0$ (or $|\langle\vec{\Xi}\rangle|^2$).

The conductance through the dot is given by
\begin{eqnarray}
G & = & \frac{e^2}{h} (2\pi\nu)^2 \sum_{i,j,\sigma,\sigma'} \left.
|\langle R, k' \sigma', j | \hat{T} | L, k \sigma, i \rangle |^2
\right|_{\varepsilon_k=\varepsilon_{k'}=\mu} \nonumber \\
 & = & \frac{e^2}{h} (2\pi\nu)^2 \sum_{i,j,\sigma,\sigma'} \left.
 \frac{\Gamma_R^j}{\Gamma_L^j+\Gamma_R^j}
 \frac{\Gamma_L^i}{\Gamma_L^i+\Gamma_R^i}
|\langle \psi_{k' \sigma'}^{(j)} | \hat{T} | \psi_{k \sigma}^{(i)}
\rangle |^2 \right|_{\varepsilon_k=\varepsilon_{k'}=\mu}
\end{eqnarray}
where $\Gamma_{\alpha}^i=\pi\nu |V_{\alpha,i}|^2$ and
$| \psi_{k \sigma}^{(i)} \rangle =(V_{L,i} |L, k \sigma,i \rangle+
V_{R,i} |R, k \sigma, i \rangle)/V_i$.
The T-matrix can be evaluated, using the Green function for the 
pseudo-fermions, $G_{SM}(\varepsilon)=[\varepsilon-\tilde{E}_{SM}+{\rm i}
\tilde{\Delta}_{SM}]^{-1}$, as in Appendix A. This yields
Eq.\ (\ref{eq:cond2}) in the text.

%---------------------------------------------

%---------------------------------------------
\pagebreak

\noindent
{\large Figure captions}
\vspace{.5cm}

Fig.\ 1:
(a) The energy diagram for the spin states $|SM\rangle$
considered in our model.
$\Delta=E_{00}-E_{S=1}$ and $E_{\rm Z}$ is the Zeeman splitting.
(b) Spin flip processes between the spin states.
The exchange couplings $J^{(i)}$ involving the spin-triplet state only
are accompanied by the scattering of conduction electrons of channel $i$.
Those involving both the spin-triplet and singlet states, $\tilde{J}$,
are accompanied by the interchannel scattering of conduction electrons.

\vspace{0.5cm}

Fig.\ 2:
The scaling calculations of the Kondo temperature $T_{\rm K}$
as a function of $\Delta$, (a) when the Zeeman splitting is 
irrelevant, $E_{\rm Z} \ll T_{\rm K}$, and (b) in a case of
$E_{\rm Z} = -\Delta$.
$D_0$ is the bandwidth in the leads.
In both the figures,
{\it a}, $\theta/\pi=0.25$; {\it b}, $0.15$; and {\it c}, $0.10$
where $\tan^2\theta=J^{(2)}/J^{(1)}$. 

\vspace{0.5cm}

Fig.\ 3:
The mean field calculations for the Kondo effect in a quantum dot
with $S=1/2$.
(a) The Kondo temperature $T_{\rm K}$ and (b) conductance through the
dot, $G$, as functions of the Zeeman splitting $E_{\rm Z}$.
$T_{\rm K}$ and $E_{\rm Z}$ are in units of
$D_0 \exp (-1/2\nu J)$ and $G$ is in
units of $(2e^2/h) \cdot 4\Gamma_L\Gamma_R/(\Gamma_L+\Gamma_R)^2$.
Inset in (a): The Kondo resonant states created around the Fermi
level $\mu$ in the leads, (A) in the absence and (B) presence of
the Zeeman splitting. The resonant width is given by $T_{\rm K}$.

\vspace{0.5cm}

Fig.\ 4:
The mean field calculations for the Kondo effect in the
present model. The Zeeman splitting is disregarded
($E_{\rm Z} \ll T_{\rm K}$).
(a) The Kondo temperature $T_{\rm K}$ and (b) conductance through the
dot, $G$, as functions of $\Delta=E_{00}-E_{S=1}$.
$T_{\rm K}$ and $\Delta$ are in units of
$D_0 \exp (-1/\nu J_{\rm MF})$. $G$, in
units of $2e^2/h$, is evaluated in a symmetric case of
$\Gamma_L^{i}=\Gamma_R^{i}$ ($i=1,2$).
$\displaystyle \tan\varphi=\sqrt{3}\tilde{J}/J_{\rm MF}$ where
{\it a}, $\varphi/\pi=0.25$; {\it b}, $0.15$; and {\it c}, $0.10$.
Note that $\varphi/\pi \le 1/6$ in this approximation (case {\it a}
is only for reference).
Inset in (a): The Kondo resonant states for $S=1$ (solid line)
and for $S=0$ (dotted line) when
(A) $\Delta \gg T_{\rm K}(0)$, (B) $\Delta \sim T_{\rm K}(0)$, and (C)
$\Delta <0$.

\vspace{0.5cm}

Fig.\ 5:
The mean field calculations for the Kondo temperature
$T_{\rm K}$ in the present model. $\varphi/\pi=0.15$ where
$\displaystyle \tan\varphi=\sqrt{3}\tilde{J}/J_{\rm MF}$.
All of $E_{\rm Z}$, $\Delta$, and $T_{\rm K}$
are in units of $D_0 \exp (-1/\nu J_{\rm MF})$.
(a) $T_{\rm K}$ is plotted in $E_{\rm Z}$-$\Delta$ plane, by contour
lines drawn every $0.25$.
The brighter color indicates the larger values of $T_{\rm K}$.
(b) $T_{\rm K}$ as a function of $\Delta$ when $E_{\rm Z}$ is fixed
at {\it a}, $0$; {\it b}, $1$; {\it c}, $2$; 
{\it d}, $5$; and {\it e}, $10$.
The broken line indicates $T_{\rm K}$ in the case of $-\Delta=E_{\rm Z}$.

\vspace{0.5cm}

Fig.\ 6:
The mean field calculations for the conductance $G$
in the present model. $G$ is plotted in $E_{\rm Z}$-$\Delta$ plane,
by contour lines drawn every $0.2 \times (2e^2/h)$.
The brighter color indicates the larger values of $G$.
$E_{\rm Z}$ and $\Delta$ are in units of $D_0 \exp (-1/\nu J_{\rm MF})$.
$\varphi/\pi=0.15$ where
$\displaystyle \tan\varphi=\sqrt{3}\tilde{J}/J_{\rm MF}$, and
$\Gamma_L^{i}=\Gamma_R^{i}$ ($i=1,2$).

%---------------------------------------------
\pagebreak

% --------------- Fig.1--------------
\vspace*{2cm}
  \epsfig{file=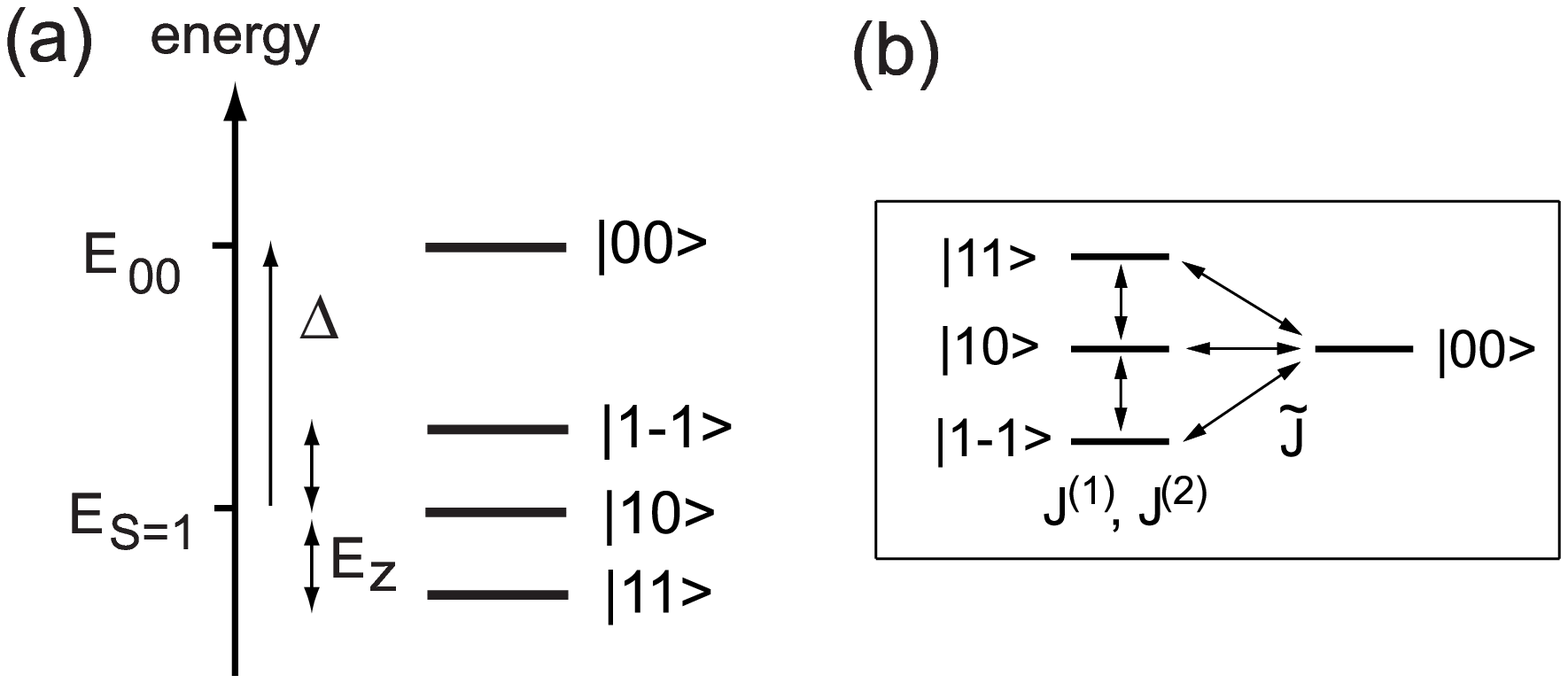,width=10cm}

  \vspace*{1cm}

\large{Figure 1 (M.\ Eto and Yu.\ V.\ Nazarov)}

%---------
\pagebreak

% --------------- Fig.2--------------
  \epsfig{file=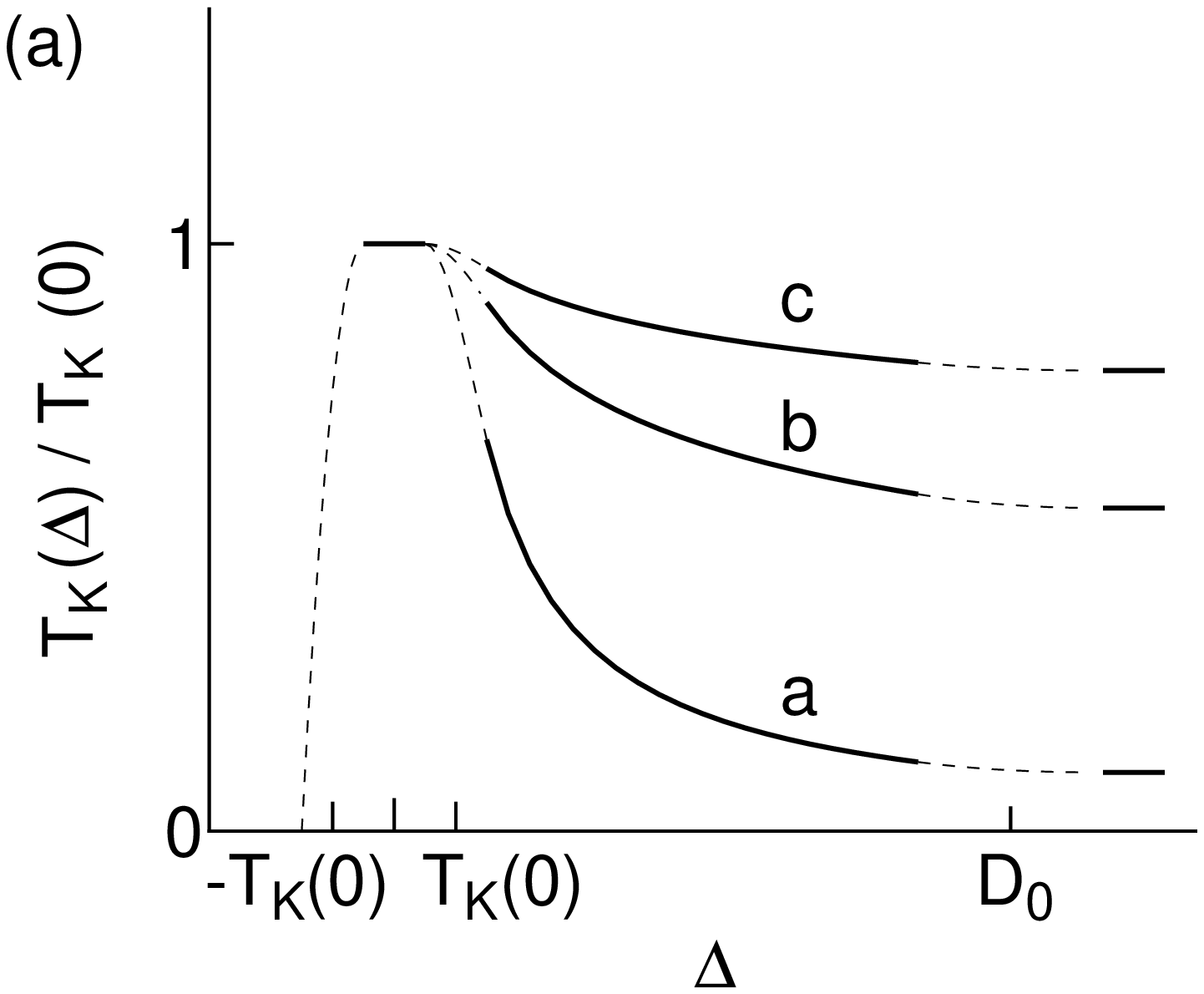,width=10cm}
  \vspace{-0.5cm}

  \epsfig{file=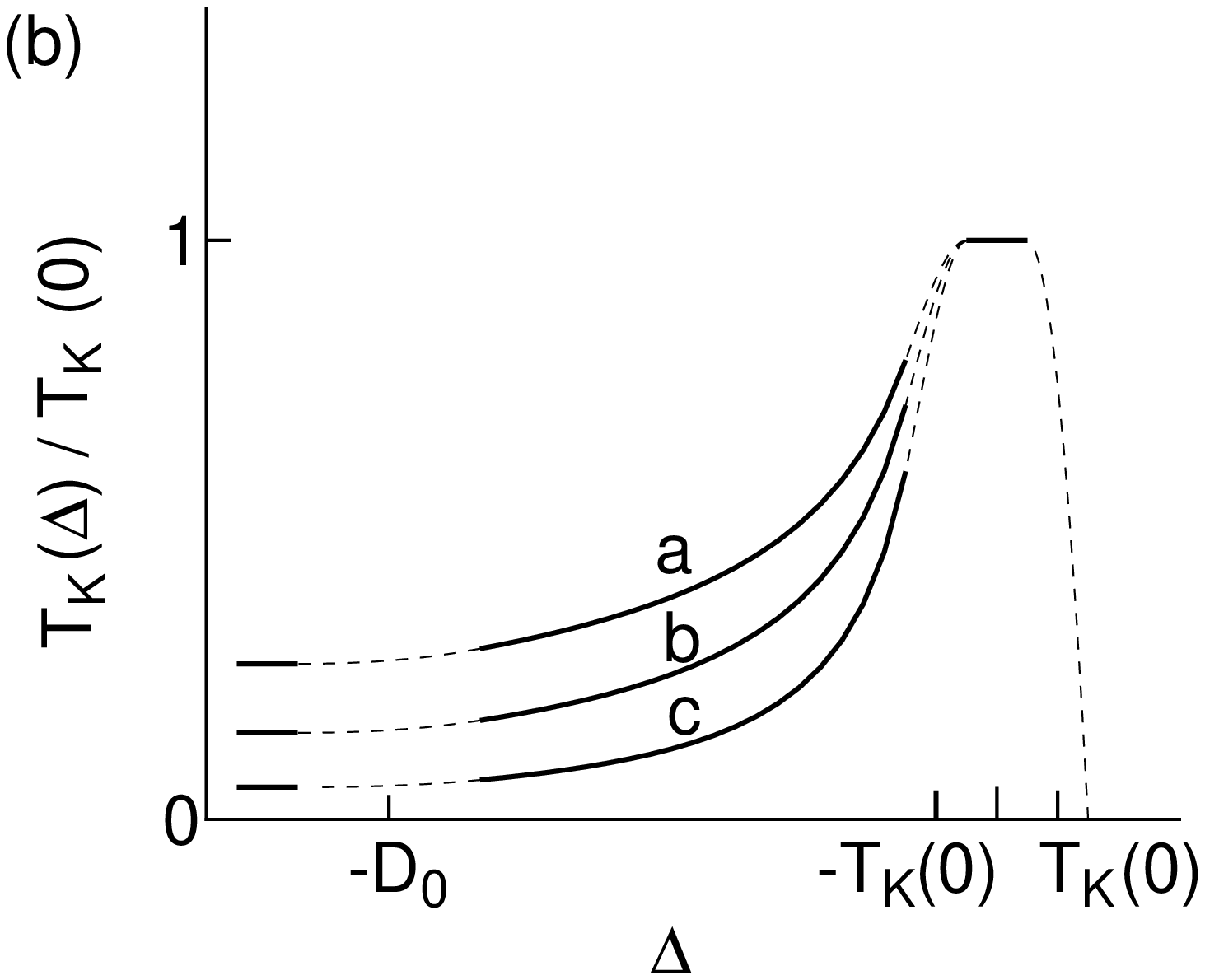,width=10cm}

  \vspace*{1cm}

\large{Figure 2 (M.\ Eto and Yu.\ V.\ Nazarov)}

%---------
\pagebreak

% --------------- Fig.3--------------
  \epsfig{file=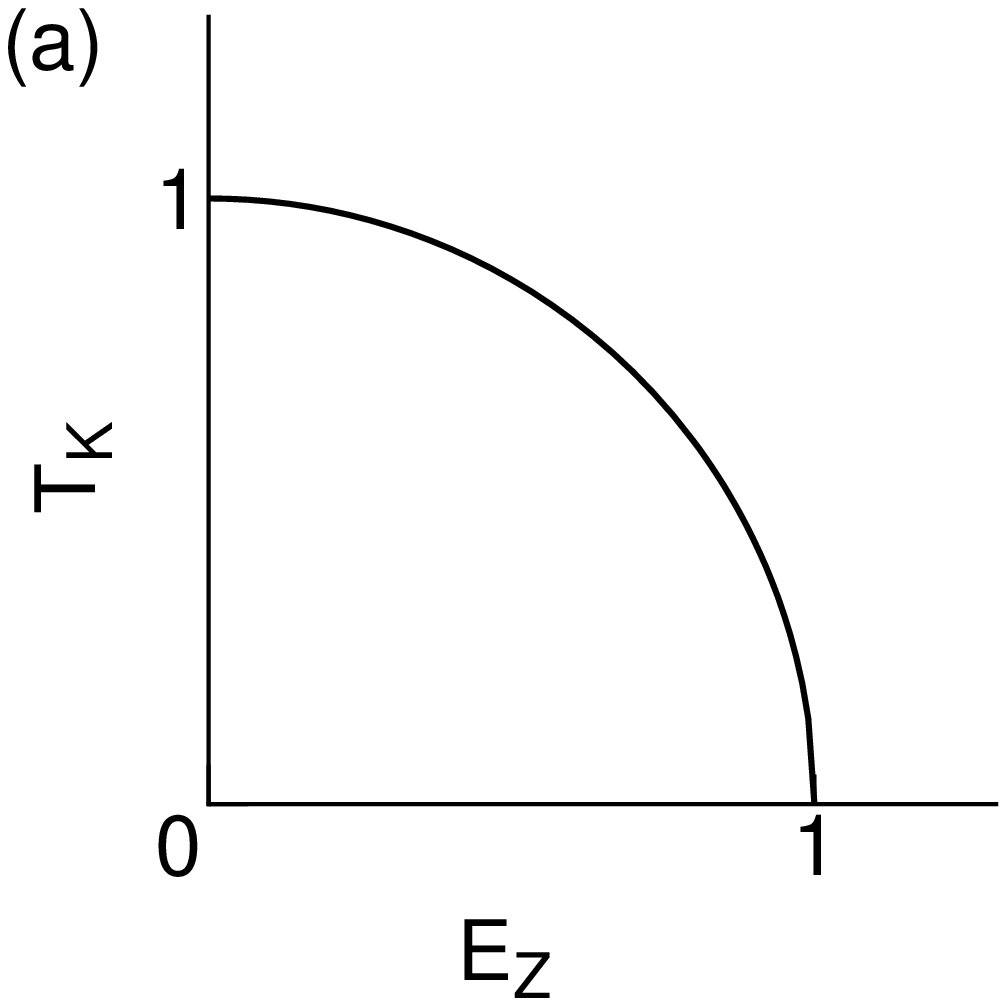,width=7cm}
  \hspace{-4cm}
  \epsfig{file=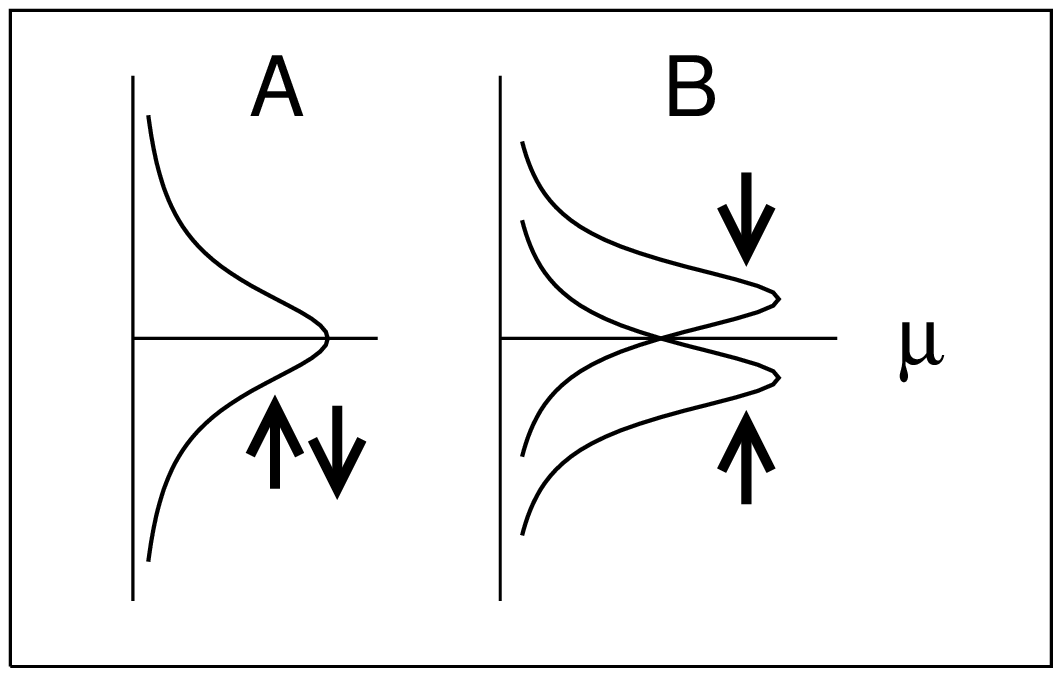,width=3.5cm}
%                 the lower limit of y-axis is changed from
%                 420 to 220 to make space below the figure.
  \hspace{-.8cm}
  \epsfig{file=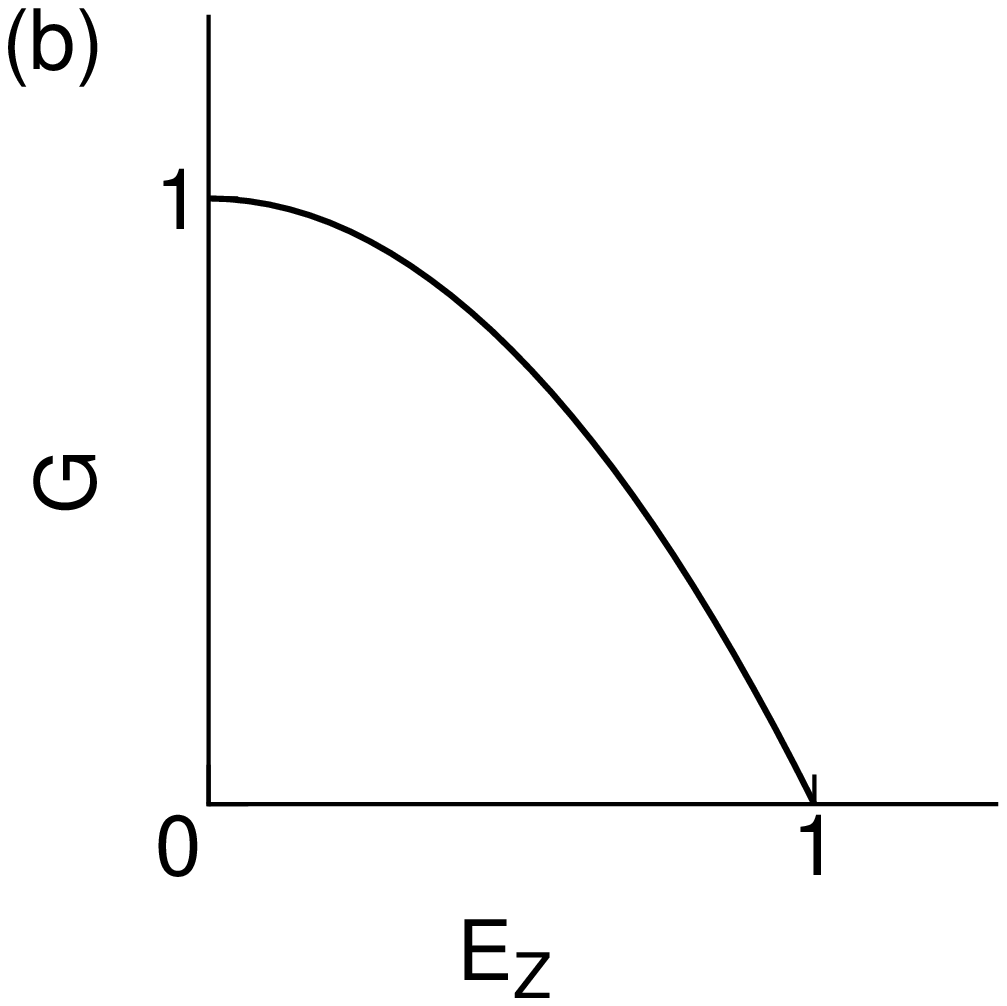,width=7cm}

  \vspace*{1cm}

\large{Figure 3 (M.\ Eto and Yu.\ V.\ Nazarov)}

%---------
\pagebreak

% --------------- Fig.4--------------
  \epsfig{file=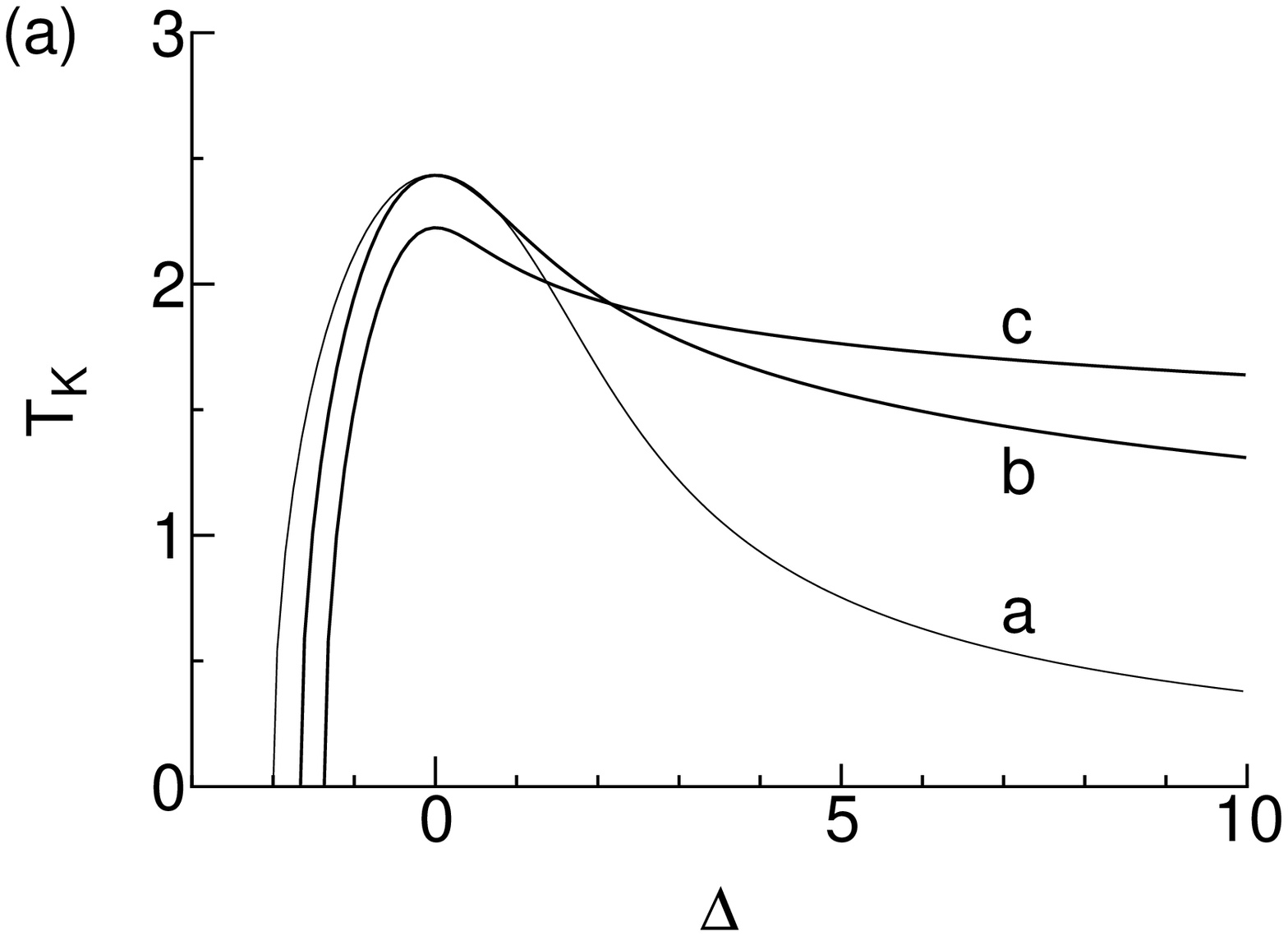,width=10cm}
  \hspace{-5.5cm}
  \epsfig{file=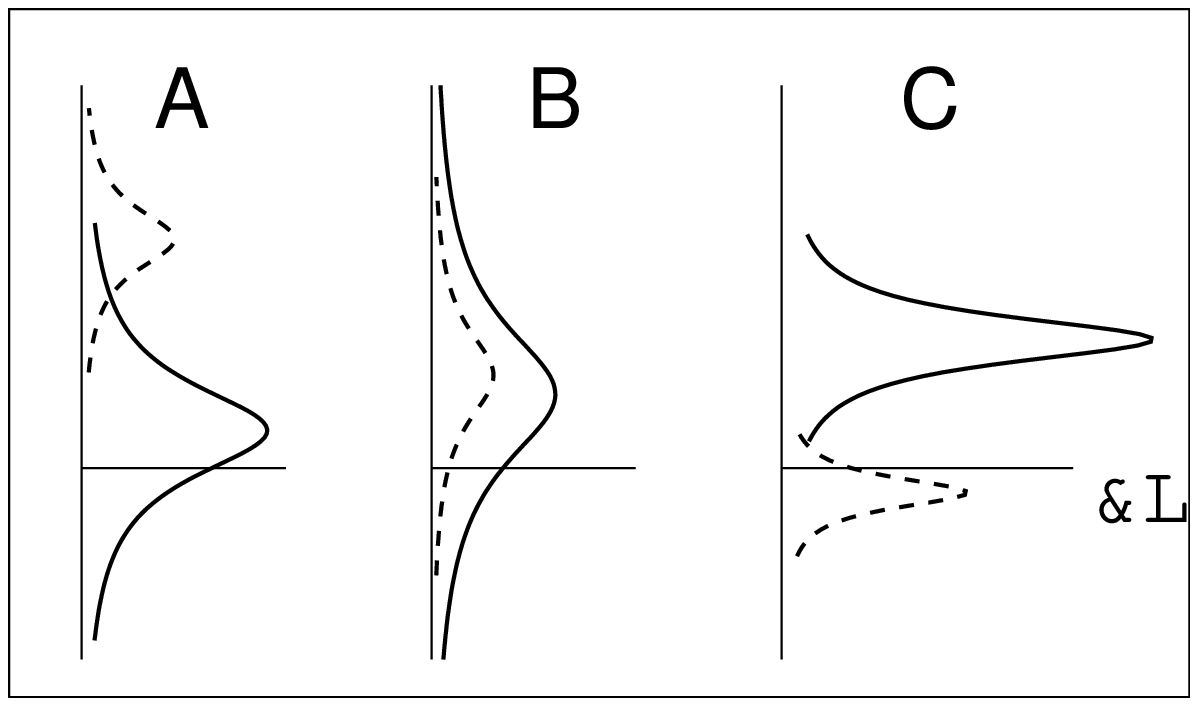,width=4.5cm}
%                 the lower limit of y-axis is changed from
%                 420 to 20 to make space below the figure.

  \epsfig{file=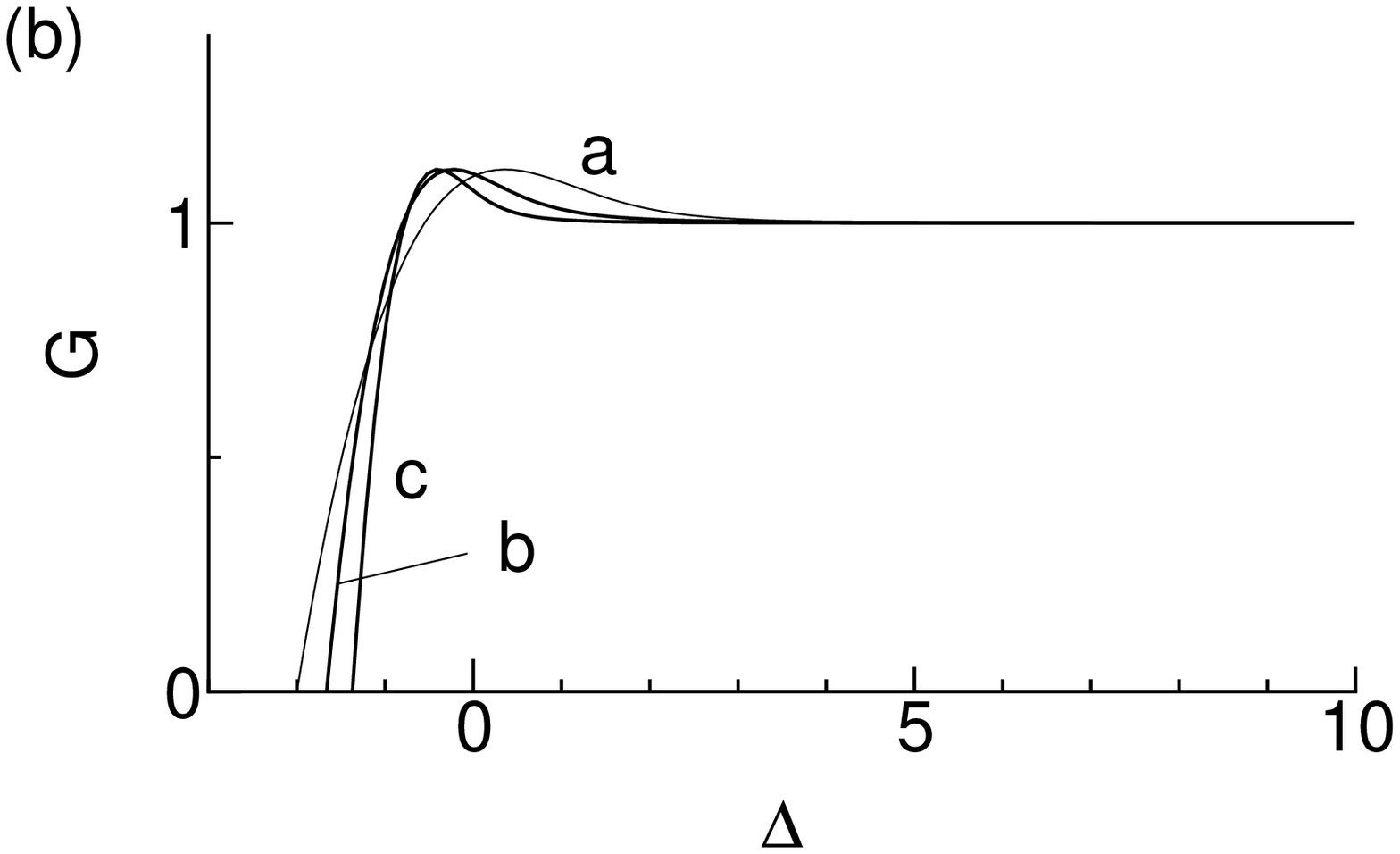,width=10cm}

  \vspace*{1cm}

\large{Figure 4 (M.\ Eto and Yu.\ V.\ Nazarov)}

%---------
\pagebreak

% --------------- Fig.5--------------
  \epsfig{file=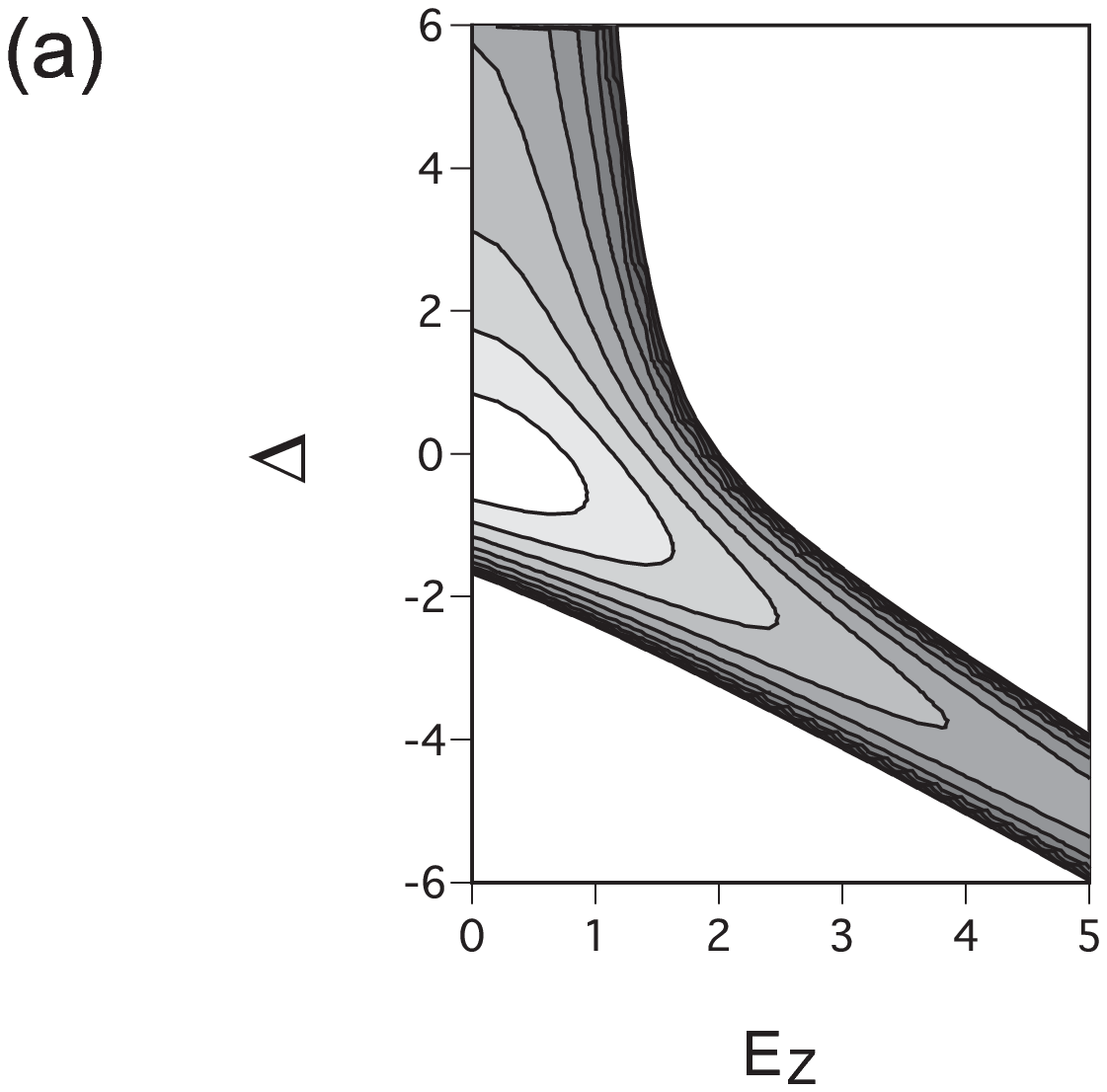,width=9cm}

  \epsfig{file=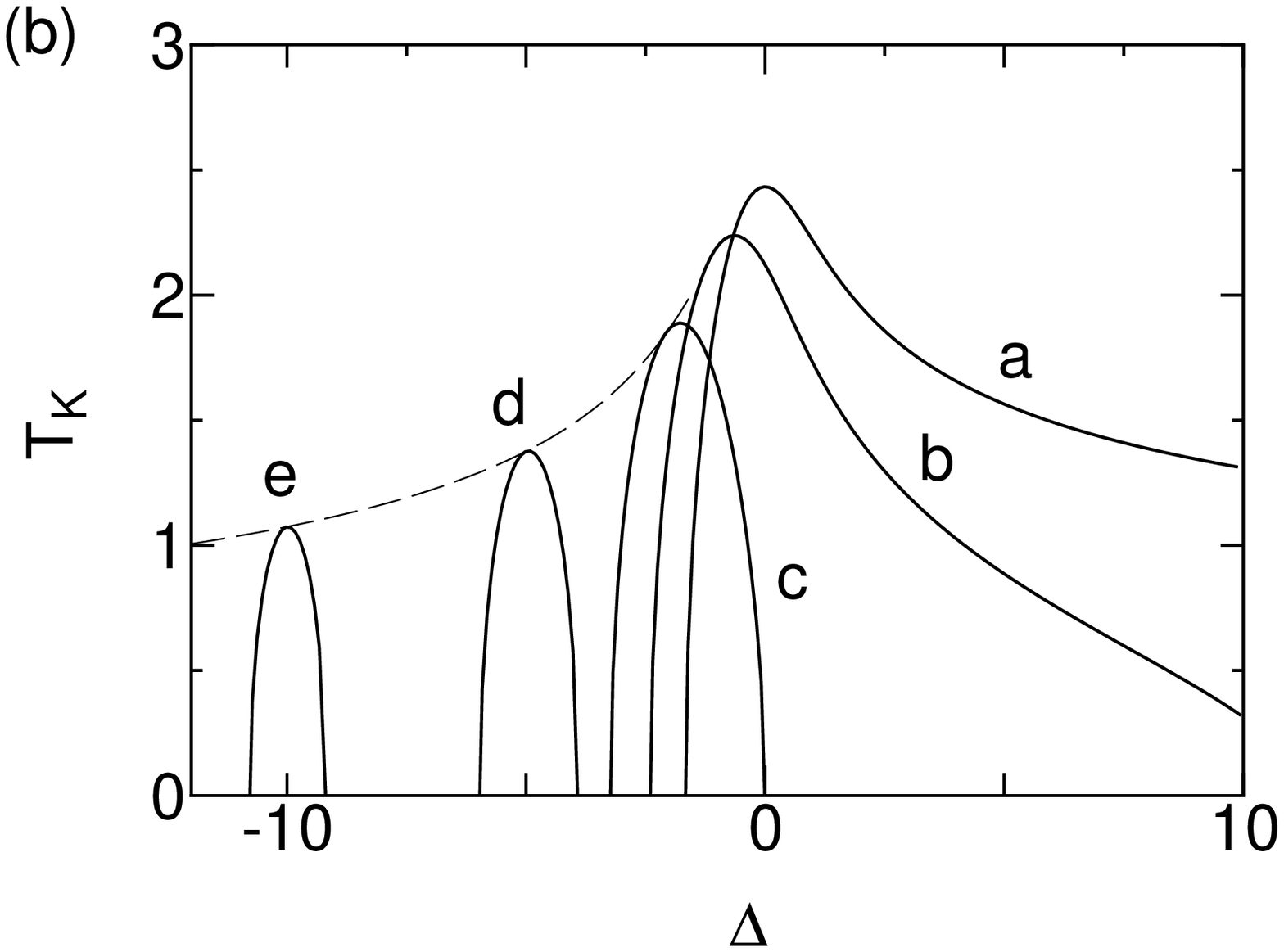,width=10cm}

  \vspace*{1cm}

\large{Figure 5 (M.\ Eto and Yu.\ V.\ Nazarov)}

%---------
\pagebreak

% --------------- Fig.6--------------
  \epsfig{file=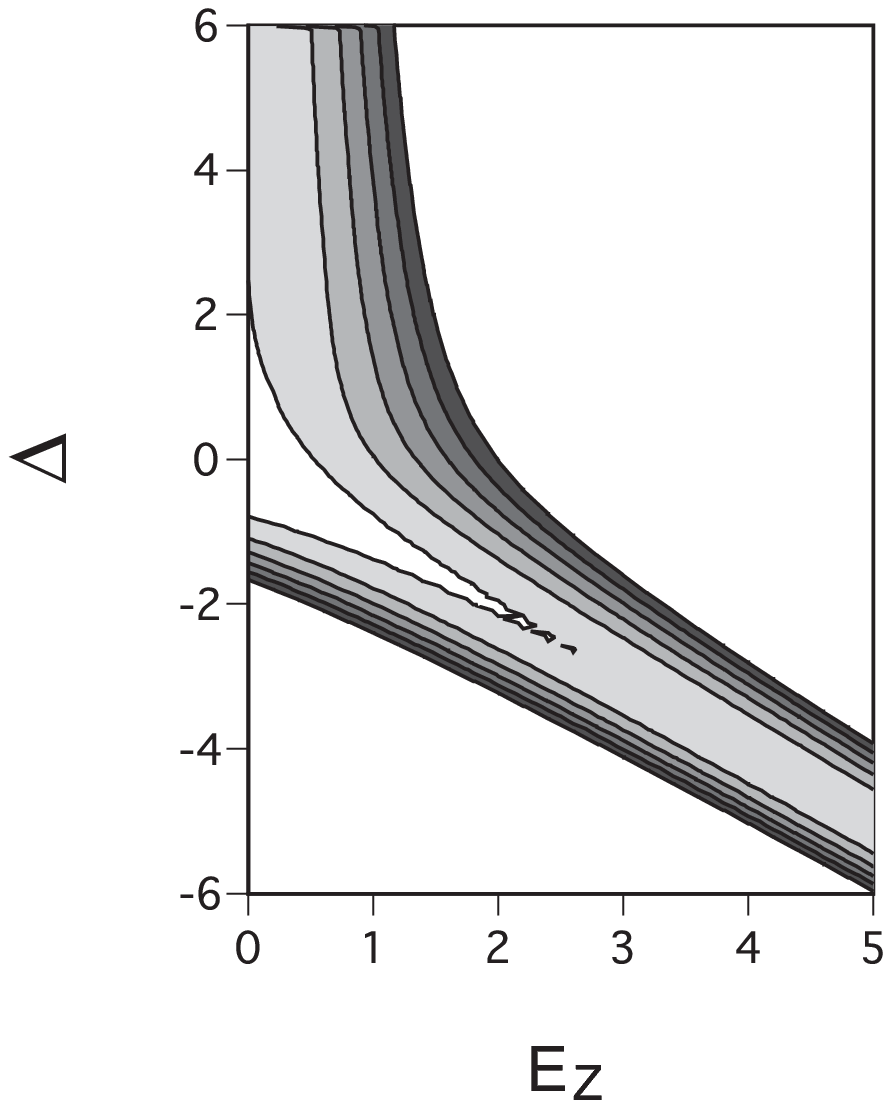,width=9cm}

  \vspace*{1cm}

\large{Figure 6 (M.\ Eto and Yu.\ V.\ Nazarov)}


\begin{references}
\bibitem{com0}
This mean field theory is equivalent to the mean field theory using
the slave bosons for $U=\infty$ Anderson model in the Kondo
region.\cite{classic2}
This method is exact in the large $N$ limit
when the dot state is $N$-fold degenerate.
\bibitem{symmetry}
The different symmetry means different orbital quantum number for
the one-electron states in a quantum dot.
We consider the situation for vertical quantum dots where the
orbital quantum number $i$ is conserved in the tunneling processes
between the dot and leads.
\bibitem{com1}
One finds that $C_1=C_2=1$ in the case of $\delta=0$, considering
the matrix element of the Coulomb interaction between
$d_{1 \uparrow}^{\dagger} d_{1 \downarrow}^{\dagger} |0 \rangle$
and $d_{2 \uparrow}^{\dagger} d_{2 \downarrow}^{\dagger} |0 \rangle$,
$\langle 22|e^2/r|11 \rangle$. This matrix element is not zero in vertical
dots of rectangular shape.\cite{Sasaki} It could disappear in circular
dots when the total angular momentum is different between the
electronic states.
\end{references}
\end{document}